# Parallel Guanine Duplex and Cytosine Duplex DNA with Uninterrupted Spines of $Ag^{I}$-Mediated Base Pairs

Steven M. Swasey[1§], Frédéric Rosu[2§], Stacy M. Copp[3], Valérie Gabelica[4*] and Elisabeth G. Gwinn[5*]

[1] Department of Chemistry and Biochemistry, UCSB, Santa Barbara, CA 93117, USA

[2] Institut Européen de Chimie et Biologie, Université de Bordeaux, CNRS & Inserm (IECB, UMS3033, US001), 2 rue Robert Escarpit, 33607 Pessac, France.

[3] Center for Integrated Nanotechnologies, Los Alamos National Laboratories, Los Alamos, NM 87545, USA.

[4] Laboratoire Acides Nucléiques: Régulations Naturelle et Artificielle, Université de Bordeaux, Inserm & CNRS (ARNA, U1212, UMR5320), IECB, 2 rue Robert Escarpit, 33607 Pessac, France.

[5] Department of Physics, UCSB, Santa Barbara, CA 93117, USA

§ contributed equally to this work

*Corresponding authors

**ABSTRACT:** Hydrogen bonding between nucleobases produces diverse DNA structural motifs, including canonical duplexes, guanine (G) quadruplexes and cytosine (C) i-motifs. Incorporating metal-mediated base pairs into nucleic acid structures can introduce new functionalities and enhanced stabilities. Here we demonstrate, using mass spectrometry (MS), ion mobility spectrometry (IMS) and fluorescence resonance energy transfer (FRET), that parallel-stranded structures consisting of up to 20 G–$Ag^{I}$–G contiguous base pairs are formed when natural DNA sequences are mixed with silver cations in aqueous solution. FRET indicates that duplexes formed by poly(cytosine) strands with 20 contiguous C–$Ag^{I}$–C base pairs are also parallel. Silver-mediated G-duplexes form preferentially over G-quadruplexes, and the ability of $Ag^{+}$ to convert G-quadruplexes into silver-paired duplexes may provide a new route to manipulating these biologically relevant structures. IMS indicates that G-duplexes are linear and more rigid than B-DNA. DFT calculations were used to propose structures compatible with the IMS experiments. Such inexpensive, defect-free and soluble DNA-based nanowires open new directions in the design of novel metal-mediated DNA nanotechnology.

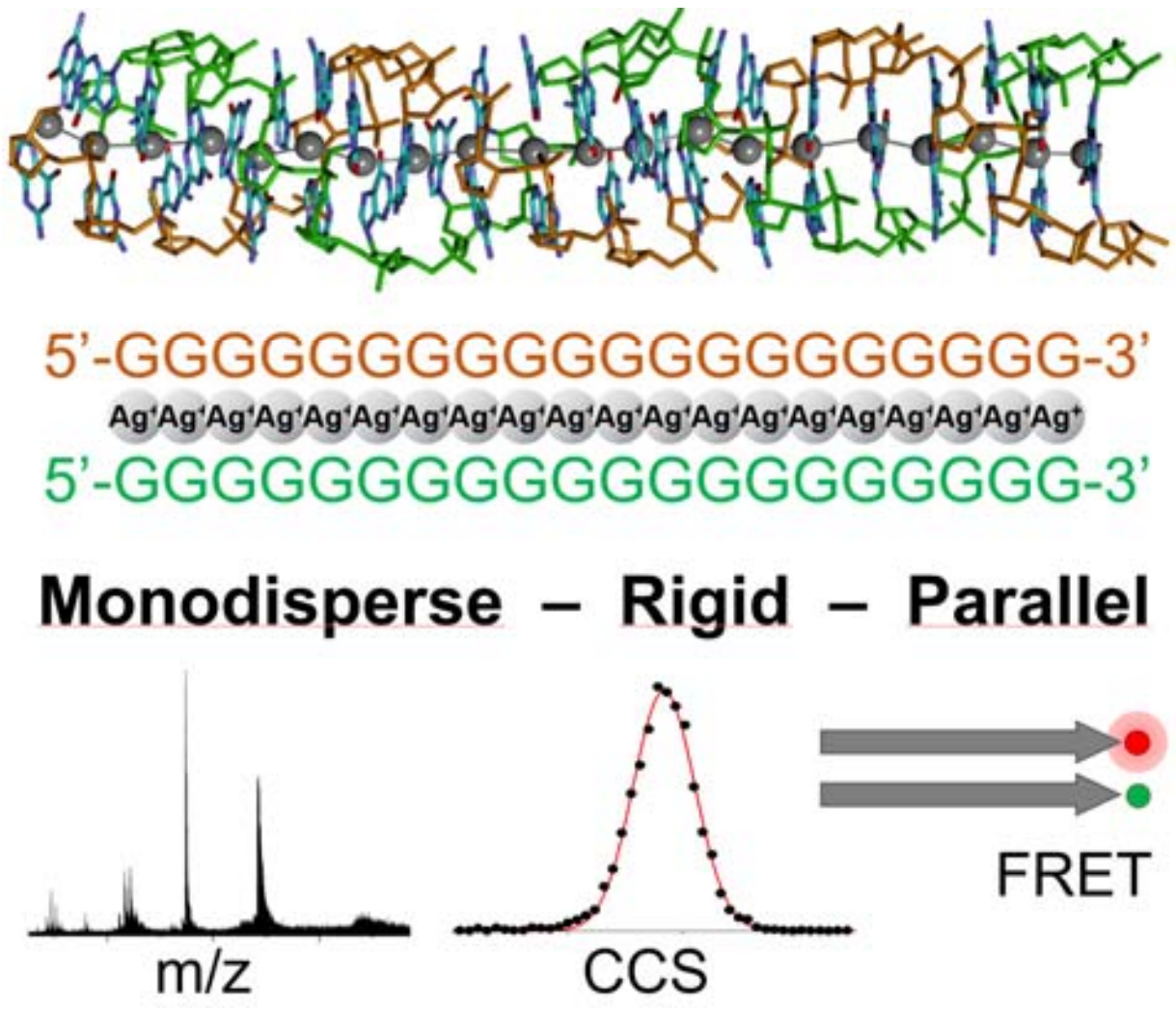

DNA is both a pivotal biological molecule and a versatile building block capable of self-assembly into intricate nanostructures. Current DNA nanotechnology is built on decades of research on the structure and thermodynamics of canonical duplex DNA. DNA also hydrogen (H) bonds to form hairpins,[1] triplex structures[2] and a variety of duplex forms in which the helicity depends on salt conditions and sequence.[3–5] Other non-canonical H-bonded forms of DNA include tetrahelical (guanine) G-quadruplex structures[6] and cytosine i-motifs.[7] Besides nanotechnology applications, these structures are suggested to have multiple regulatory functions in living organisms,[8] and were recently visualized within the DNA of human cells.[9,10]

The knowledge base about H-bonded DNA building blocks has been essential to develop efficient and ordered hybridization of single-stranded DNA into nanostructures, ranging from elaborate shapes formed from DNA duplex segments[11] to dynamic nanomachines that use non-canonical H-bonded DNA motifs for actuation in molecular robotics.[12] A complementary area of research with relevance to both the biological and structural material aspects of DNA is metal-base interactions, using either natural or artificial bases.[13–15] Many metal cations interact with various parts of DNA, but only $Hg^{2+}$, $Ag^{+}$ and $Pt^{2+}$ specifically interact with the natural bases,[16] giving them the capability to target specific base motifs in DNA, alter its biological function, or form potentially functional metallic DNA.[14,17–19] $Pt^{2+}$ complexes bind preferentially to G bases, form crosslinks in duplex DNA, interfere with cell replication and are extensively studied for anti-cancer activity.[19,20] More recently, $Hg^{2+}$ and $Ag^{+}$ were found to form T–$Hg^{II}$–T and C–$Ag^{I}$–C pairs at cytosine or thymine mismatches in antiparallel B-DNA.[21,22]

Although all such metallobase pairs could be exploited to build new DNA-based assemblies, $Ag^{+}$ is particularly intriguing because it has little known toxicity in humans and is commonly used as an antimicrobial agent.[23,24] Silver mediated base pairing has been used to control the size and optical properties of fluorescent DNA-templated silver clusters[25,26] and to build antiparallel mixed-base duplexes of $Ag^{I}$-mediated base pairs.[27] To date, however, the use of $Ag^{I}$-paired DNA in nanotechnology has been hindered by lack of knowledge of the basic properties of silver-paired structural motifs. For example, the RNA strand r(GGACU[$^{Br}$C]GACUCC) crystallized into the expected antiparallel duplex with two C–$Ag^{I}$–C base pairs,[28] but the corresponding DNA strand crystallized into slipped-stranded antiparallel structures with adenines bulged out, forming a nanowire of silver cations positioned by C–$Ag^{I}$–C, G–$Ag^{I}$–G, G–$Ag^{I}$–C and T–$Ag^{I}$–T base pairs.[27] This illustrates the difficulty to predict and control the incorporation of silver-mediated base pairs into DNA structures. Other designs with unnatural bases as coordination sites[29] have been conceived, but were also almost exclusively based on the antiparallel B-DNA template.

Recently we discovered that homo-guanine as well as homo-cytosine strands form robust bimolecular DNA structures incorporating one silver ion per base pair.[30,31] The stabilities of these $Ag^{+}$-paired strands significantly eclipse Watson-Crick (WC) paired duplexes of similar length.[30,31] Due to the ease of formation, stability and low polydispersity of products observed for short homobase oligomers, such $Ag^{+}$-mediated base pairs are particularly promising to form more robust and tunable metal-DNA nanostructures.[31] Here we studied the structures formed upon annealing $dG_n$ (n = 6–20) or $dC_n$ (n = 6–30) in $AgNO_3$ (1 equivalent per base), in the presence of 50 mM $NH_4OAc$ to ensure sufficient ionic strength. All electrospray mass spectra confirm that the major product is a duplex with exactly 1 $Ag^{+}$ per base pair. Figure 1 shows the spectra of the longest structures, with 20 G–$Ag^{I}$–G and 30 C–$Ag^{I}$–C base pairs (spectra of all other strands are shown in Figs. S1-S5). The present paper focuses on elucidating the topology and shape of these silver-mediated G-duplexes and C-duplexes.

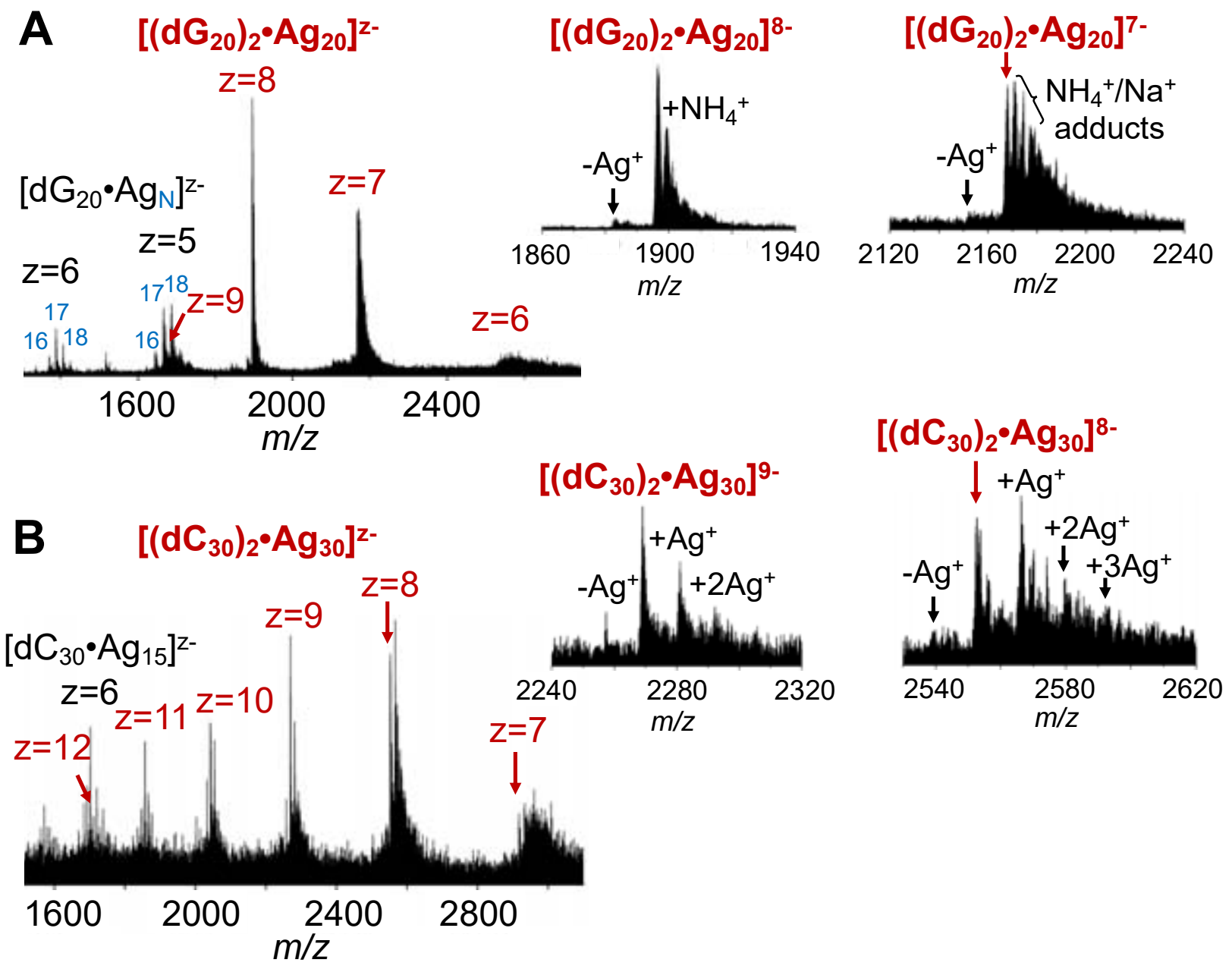

FIGURE 1. Electrospray mass spectra of the G-duplex $[dG_{20}]_2$•$Ag_{20}$ (A) and C-duplex $[dC_{30}]_2$•$Ag_{30}$ (B). The concentrations were 80 μM DNA strands at 1 equivalent $AgNO_3$ per base in 50 mM $NH_4OAc$. Top side panel: Enlargement of the major product peak in (A) shows nearly monodisperse $[dG_{20}]_2$•$Ag_{20}$ products. Lower side panel: Enlargement of the major product peak in (B) shows detectable $[dC_{30}]_2$•$Ag_n$ products with n = 29, 31, 32 and 33, in addition to n = 30.

We used fluorescence resonance energy transfer (FRET) to investigate the strand orientation in dye-labelled strands. In FRET, an excited donor dye D transfers energy non-radiatively to an acceptor dye A, with an efficiency $E = 1/(1+(R/R_0)^6)$. $R$ is the D-A separation and $R_0$, the Förster radius, is set by the dyes used in the D-A pair. Owing to the steep $R$-dependence of the efficiency, FRET is sensitive to the proximity of the donor and acceptor in the duplex. We used the D-A dye pair Alexa 488-Alexa 647 ($R_0$ = 5.2 nm), attached to the DNA by $C_6$ linkers. We selected 450 nm as the excitation wavelength so that only the Alexa 488 donor is directly excited. Thus emission from the Alexa 647 acceptor arises solely from FRET (see Section S2.1 for details). We chose a FRET scheme wherein the strand **A3′**, carrying the acceptor dye at its 3′ end, is silver-paired either with the strand **D3′** (donor dye at 3′ end) or with the strand **D5′** (donor dye at 5′ end). If strands A and D form a *parallel* duplex, **D3′•A3′** will hold the dyes at the same duplex end, giving a larger FRET signal (quenched emission from D and FRET-activated emission from A) than **D5′•A3′**, which for parallel strand orientation holds the dyes on opposite duplex ends. If *antiparallel* duplexes instead form, the FRET signal will instead be small for **D3′•A3′** (dyes on opposite duplex ends for antiparallel pairing) and large for **D5′•A3′**.

The results are shown in Figure 2. Full sequences are given in the legend. The number of bases in the central stretches of the strands ($C_{20}$ and $G_{15}$) was limited by the numerous truncated byproducts of these long homobase runs in the as-received strands, which necessitated pre-purification stages prior to FRET experiments (details in Supporting Information) and made obtaining sufficient yields of the full length, dye-labelled products quite challenging. For much shorter strands, the difference in FRET signal between parallel and antiparallel alignment would not be large enough to be definitive. Thus for FRET experiments we used only $C_{20}$ and $G_{15}$ DNA, with thymine (T) extensions on both ends to reduce dye interactions with G and C bases and promote free dye rotation. We used different lengths of the T extensions on donor and acceptor strands to enable isolation of the desired D•A duplexes by HPLC. All FRET data shown here are for these purified D•A products (see supporting information section S2.2). Silver nitrate was added at 2.1 equivalent $Ag^+$ per expected GG or CC base pair. The cytosine strands were first prepared in 500 mM ammonium acetate ($NH_4OAc$) to increase the yield of $Ag^+$-paired D•A products, because A•A and D•D byproducts predominated at low buffer concentrations.

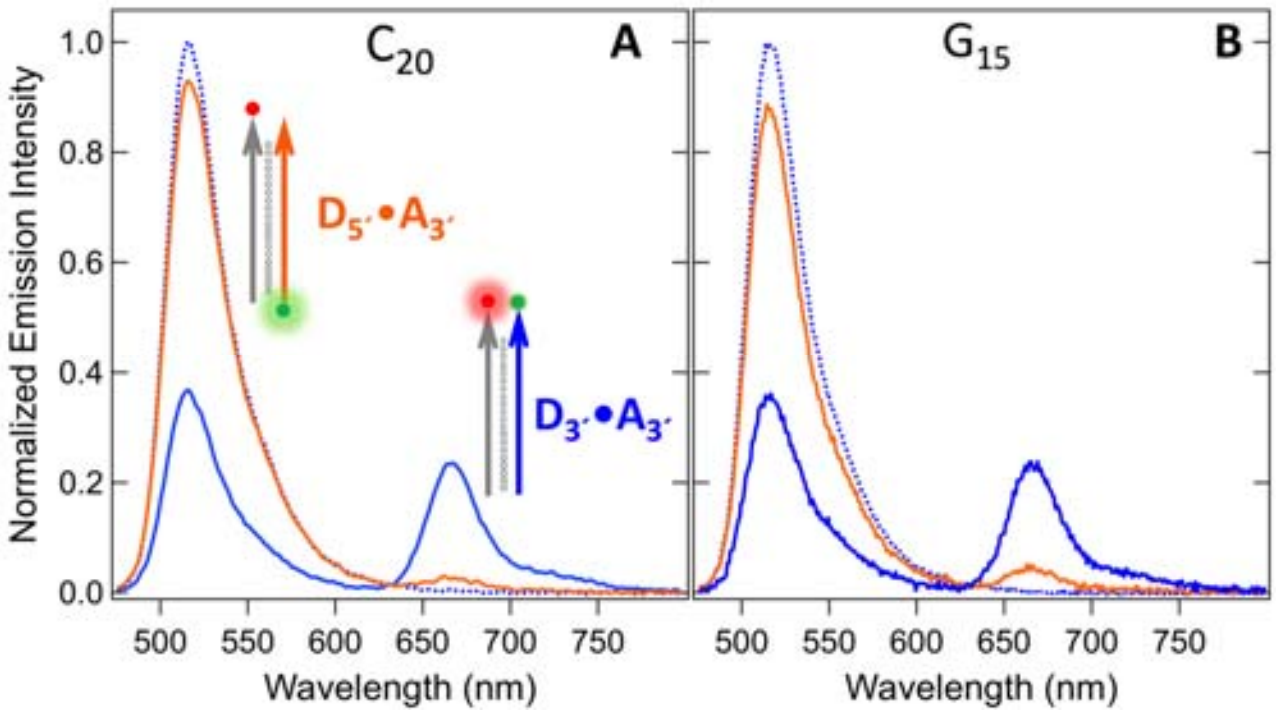

FIGURE 2. (A) Emission spectra for the cytosine case: **A3′** is 5′-d($T_5C_{20}T_5$)-[Alexa 647]-3′, **D3′** is 5′-d($T_2C_{20}T_2$)-[Alexa 488]-3′ and **D5′** is 5′-[Alexa 488]-d($T_2C_{20}T_2$)-3′. (B) Emission spectra for the guanine case: **A3′** is 5′-d($T_4G_{15}T_4$)-[Alexa 647]-3′, **D3′** is 5′-d($T_2G_{15}T_2$)-[Alexa 488]-3′ and **D5′** is 5′-[Alexa 488]-d($T_2G_{15}T_2$)-3′. Dotted blue curve is **D3′** without $Ag^+$. Solid blue curve is the purified $Ag^+$-paired **D3′•A3′**. Solid orange curve is the purified $Ag^+$-paired **D5′•A3′**. In all cases excitation is at 450 nm, which directly excites only the donor.

The HPLC-purified cytosine D•A products were solvent-exchanged into 10 mM $NH_4OAc$ for FRET analysis. The preparation of the guanine D•A heteroduplexes required first disrupting the polydisperse G-quadruplex aggregates present in the received material (details in Section S2.2.2). This was achieved by (i) adding $Ag^+$ to convert strand aggregates to $Ag^+$-paired duplexes, (ii) purifying these $Ag^+$-paired homodimers by HPLC, then (iii) removing the $Ag^+$ by chelation with cysteine. These disaggregated, single strands were then mixed with $Ag^+$ and HPLC separation was used to isolate the guanine D•A pairs for FRET measurements. The discovery that $Ag^+$ can break up multi-stranded guanine aggregates offers an interesting alternative to the use of LiOH for disrupting pre-formed G-quadruplexes[32] and suggests the possibility of employing $Ag^+$ to perturb G-quadruplexes in cells.

Figure 2 shows the measured emission spectra of the HPLC-purified, $Ag^+$-paired **D3′•A3′** duplex (solid blue curve), **D5′•A3′** duplex (orange), together with the **D3′** cytosine control strand (blue dotted curve). The spectra are normalized using the ratiometrically calculated FRET efficiencies found from the donor and acceptor emission channels (details in Section S2.3). Electrospray-ion-ization mass-spectrometry (ESI-MS) of the purified D•A complexes showed narrow $Ag^+$ distributions with highest abundance at 18 $Ag^+$ and 19 $Ag^+$ for the cytosine case (Figure S8) and at 15 $Ag^+$ for the guanine case (Figure S11), corresponding to one $Ag^+$ per G-G base pair and very close to one $Ag^+$ per C-C base pair. The C-duplexes (Fig. 2A) and G-duplexes (Fig. 2B) exhibit the same behavior: **D3′•A3′** shows a large FRET signal with strong quenching of donor emission and activation of acceptor emission, whereas **D5′•A3′** shows only slight donor quenching and very little acceptor emission. (For the **D3′•A3′** duplexes, which hold the D and A dyes on the same end of the duplex, the FRET efficiencies and the relation $E = 1/(1+(R/R_0)^6$ give a separation of 4.7 nm between the dyes, consistent with the duplex width plus the additional dye separation that arises from the thymine extensions and alkyl dye linkers. For the **D5′•A3′** duplexes, which hold the dyes on opposite duplex ends, the FRET efficiencies give separations of 7.2 nm for the guanine case and 8.0 nm for the cytosine case.) These FRET signatures thus indicate that the $Ag^+$-paired C-duplexes and G-duplexes are parallel.

For cytosine-rich strands, a parallel duplex was previously suggested for much shorter, 8-base cytosine DNA strands[33] based on indirect evidence. Calculations for $Ag^+$ pairing of model $[dC_n]\bullet Ag_n$ duplexes also concluded that parallel arrangements are more stable than antiparallel ones.[34] However, parallel structures are incompatible with the i-motif-like intramolecular structure proposed for 5′-$(TAACCC)_4$-3′in presence of $Ag^+$.[35] The effect of mixed base runs between short cytosine tracts thus remains to be investigated. In mixed base strands, G–$Ag^I$–G base pairs were previously observed in an antiparallel strand arrangement[27] and in a parallel arrangement,[36] again showing the sensitivity of silver paired DNA structure to sequence. In the absence of $Ag^+$, parallel-strand forms are common in G-quadruplexes containing a single stretch of guanines, but sequences with stretches of 15 or more guanines rather fold into intramolecular structures,[37] or bimolecular G-quadruplexes at low $NH_4OAc$ concentrations,[38] rather than tetramolecular G-quadruplexes with all guanines involved in consecutive G-quartets.

We used ion mobility spectrometry (IMS) to compare the overall topology of the $Ag^I$-paired C-duplexes and G-duplexes with that of antiparallel duplexes, bimolecular i-motifs and bimolecular G-quadruplexes. Ion mobility spectrometry measures the drift time of ions dragged through a buffer gas (here, helium) by an electric field.[39] The arrival time distribution is then converted in a distribution of collision cross sections[40] (CCS, in $Å^2$), which represent the surface area of the ion responsible for the ion-gas momentum transfer (more in supporting information Section S3). The experimental CCS values can then be compared with values obtained by trajectory calculations[41] on proposed structural models. One important piece of information is the evolution of the CCS with the size of the system: a linear evolution indicates growth into a rod-like structure, whereas an evolution with a power of ~2/3 indicates spherical growth.[42] Further, the width of each CCS distribution indicates the variety of structures co-existing under a given mass peak.[40] Finally, the width of the charge state distribution and the extent to which the CCS changes with the charge states indicate the flexibility of the molecule during the electrospray phenomenon.[43] We rendered all these features in the form of violin plots (Figure 3).[40] The x-axis is the number of base pairs and the y-axis is the CCS distribution, mirrored so that the center is correctly placed on the x-axis, and sized to indicate the relative abundance of each charge state.

Figure 3 shows the CCS distributions of $[dG_n]_2\bullet Ag_n$, $[dC_n]_2\bullet Ag_n$, and the Watson-Crick duplexes $[d(CG)_n]_2$. The $[dG_n]_2\bullet Ag_n$ results (Fig. 3A) are remarkable in several respects: (i) the CCS increases perfectly linearly with size, (ii) the CCS distributions are the narrowest, (iii) the charge state distributions are the narrowest (see also SI), and (iv) the CCS is little influenced by the charge state. All aspects concur to indicate that a rod-like and rigid structure persists up to the 20-mer G–

$Ag^I$–G duplex. G-quadruplex structures are rigid as well (see the results for $[dTG_nT]_2$•$(NH_4^+)_i$ in Fig. S12), but the aspect ratio differs: for identical numbers of bases, the CCS values are 15-20% larger for the $[dG_n]_2$•$Ag_n$ structures, suggesting higher length/diameter aspect ratios than the G-quadruplexes. This is compatible with a double-helical (not tetra-helical) structure.

We thus built an atomistic model for the parallel-stranded duplex structures containing contiguous G–$Ag^I$–G base pairs, computed their CCS and compared them with the experimental ones. We first built the 6-bp duplex $[dG_6]_2$•$Ag_6$ as follows. A parallel-stranded backbone with adequate spacing for two purines was extracted from the structure of parallel right-handed poly(A) RNA (PBD: 4JRD).[44] The backbone was converted from RNA to DNA and the AA base pairs were removed. The silver ion in the GG base pair was assumed to coordinate with the N7 position of guanines, as predicted in calculations[30] and found in crystallographic structures.[27] The G–$Ag^I$–G base pair geometry was taken from the PBD structure 5XJZ[36], then fit in the parallel backbone. Because experimentally the main charge state of the complex with 6 $Ag^+$ is 4-, all 10 phosphate groups were left deprotonated. The entire structure was then optimized using DFT calculation with M06-2X functional[45] including the dispersion correction GD3[46] (full details in SI section S4). Ab-initio molecular dynamics (BOMD) was then performed on the optimized structures. A representative structure is shown in Fig. 3D. The CCS values calculated using the trajectory model for 200 structures along the BOMD agree very well with the experimental values (the dots on Figure 3A shows the average values; histograms are shown in Fig. S14).

The base pairs remained mostly planar during the BOMD simulation, and no hydrogen bonds form between guanines belonging to adjacent base pairs. However, the guanine amino groups are able to make hydrogen bonds with phosphate oxygens (Figure S15). These H-bonds contribute to rigidify the structure. The optimized backbone conformation is very close to the starting structure of the parallel RNA. The average Ag—Ag distance in the five nearest neighbor pairs over the BOMD is 3.17 Å (standard deviation: 0.13 Å), which is typical of argentophilic bonds.[47] We then generated atomistic models for the longer $[dG_n]_2$•$Ag_n$ duplexes by bridging $[dG_6]_2$•$Ag_6$ subunits and removing bases to achieve the desired length (Fig. S17). The CCS values calculated in helium with the trajectory model ($^{TM}CCS_{He}$) for these longer G-duplex models also match the experimental values, suggesting that our atomistic model appropriately captures the aspect ratios of all G-duplexes up to the 20-mer. The natural poly(dG) backbone is thus a perfectly adequate scaffold to align a silver wire linked by argentophilic bonds.

The $[dC_n]_2$•$Ag_n$ duplexes showed broader CCS distributions (Fig. 3B). For the longest structures (20 to 30 base pairs) the distributions are clearly bimodal. The charge state distributions were broader as well, suggesting that the structures remained more flexible during the electrospray charging and desolvation process. We used the atomistic model for the parallel C-duplexes proposed by Lopez Acevedo[48] to generate a starting structure for $[dC_6]_2$•$Ag_6$, which was then DFT optimized (Figs. 3E and S18), subjected to BOMD, and bridged to obtain the expected trend line for rod-like structures. This trend line corresponds to the most extended conformations, which are observable only for the highest charge states. Because of the greater flexibility of the $[dC_n]_2$•$Ag_n$ duplexes, the ion mobility results alone do not confirm whether the parallel-stranded structure found in solution for $[dC_{20}]_2$•$Ag_{20}$ is maintained across the entire size range. However, the evolution is progressive from n=6 to n=30, and thus parallel solution structures are plausible for all.

Gas-phase compaction at the low charge states generated from ammonium acetate was reported previously for Watson-Crick duplexes,[49] but modeling was based on force fields. Here, we used higher-level DFT to optimize a 8-bp $[d(CGCGCGCG)]_2$ duplex (Figs. 3F and S19). In the optimized structures, the minor groove (but not the major groove) is bridged by phosphate-phosphate hydrogen bonds. The Watson-Crick base pairs and stacking between the base pairs are also preserved. We then built the trendline by concatenation of these subunits, and compare it with the experimental CCS distributions for the $[d(CG)_n]_2$ duplexes (Fig. 3C). The trend line matches the experimental CCS of the duplexes up to 18 base pairs. Starting at 20 base pairs (Fig. 3E), the distribution becomes bimodal, and the trendline matches the extended form. From 26-bp on, duplexes do not acquire enough charges to maintain extended structures. Kinks can thus form lower-energy conformations involving additional (non-native) phosphate-phosphate hydrogen bonds form in the gas phase, most probably across both grooves. Similar kinks and extra H-bonds can occur in the C-duplexes (explaining the bimodal distributions), but not in the G-duplexes wherein the guanine bases lock the phosphate groups in place.

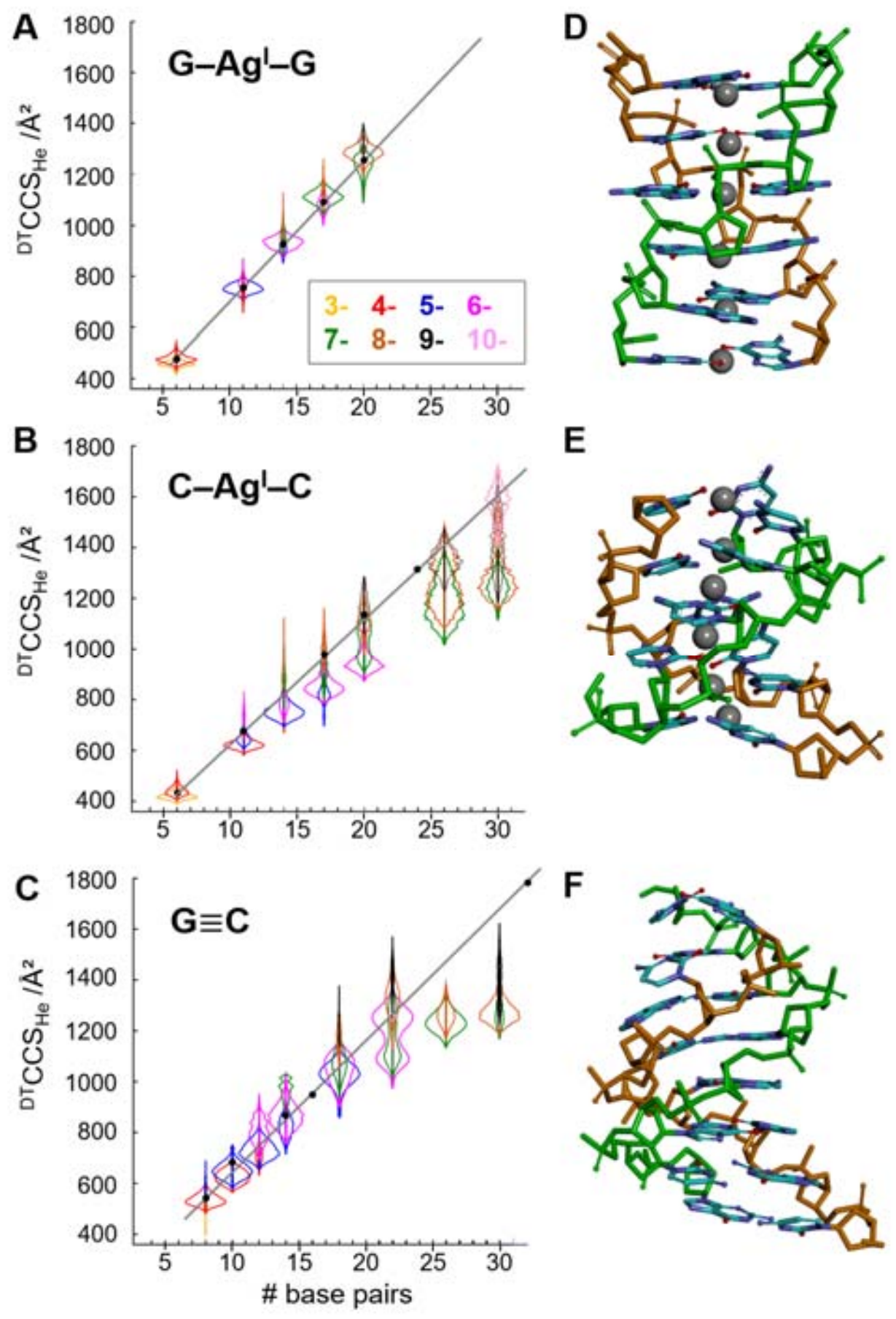

FIGURE 3. (A-C) Collision cross section distributions for (A) $[dG_n]_2•Ag_n$ duplexes, (B) $[dC_n]_2•Ag_n$ duplexes, and (C) $[d(CG)_n]_2$ duplexes, as a function of the number of base pairs. The charge states are color coded as shown in the inset. (D-F) DFT-optimized structures for a parallel $[dG_6]_2•Ag_6$ duplex (D), a parallel $[dC_6]_2•Ag_6$ duplex (E) and the antiparallel $[d(CG)_4]_2$ duplex (F). The lines in panels A-C indicate the calculated $^{TM}CCS_{He}$ values as a function of duplex size, based on these atomistic models (see SI for full details on the generation of the structures).

In conclusion, FRET experiments showed parallel-stranded duplex structures for runs of 15 sequential G-$Ag^{I}$-G pairs or 20 sequential C-$Ag^{I}$-C pairs. Ion mobility spectrometry showed very similar shape factors for all $Ag^{I}$-mediated poly(dG) duplexes up to 20 base pairs. $Ag^{I}$-mediated poly(dC) duplexes are more flexible, and the structural assignment based on collision cross sections is less straightforward, but the close agreement between the measured CCS distribution and calculations for the parallel-stranded $[dC_6]_2{\bullet}Ag_6$ duplex (Fig. S14A) suggest that a parallel duplex form is also plausible for short constructs. This information on the topology of strand pairing is essential for incorporation of the highly stable, structurally regular "wires" of sequential silver cations into designs for new DNA nanotechnology.[31] These structures are simply prepared (mixing and annealing), and further optimization to prepare even longer nanowires by circumventing intramolecular folding is certainly possible. We additionally found that $Ag^+$ converts solutions of known quadruplex-forming DNA strands into the silver-paired duplex and that subsequent chelation with cysteine can be used to remove the silver, opening new possibilities for manipulating G quadruplexes. The silver-mediated parallel G-duplexes are particularly robust and rigid, which makes them an attractive new scaffold for DNA-based nanotechnologies.

## ASSOCIATED CONTENT

### Supporting Information

Supporting Information is available free of charge on the ACS Publications website.

Supplementary materials and methods, full ESI-MS results, detailed FRET results, full IM-MS results (CCS distributions) on G-quadruplexes and i-motifs, computational details and calculated structures of longer duplexes. (PDF)

## AUTHOR INFORMATION

### Corresponding Author

Correspondence should be addressed to Elisabeth G. Gwinn (bgwinn@physics.ucsb.edu) or Valérie Gabelica (v.gabelica@iecb.u-bordeaux.fr).

### Notes

The authors declare no competing financial interests.

## ACKNOWLEDGMENT

We acknowledge support from the United States NSF (grant CHE-1213895 to EGG) and the European Research Council under the European Union's Seventh Framework Program (ERC grant 616551 to VG). The research reported here made use of shared facilities of the UCSB MRSEC (NSF DMR 1720256) and of the Plateforme de BioPhysico-Chimie Structurale at IECB.

# Supplementary Information for:

# Parallel Guanine Duplex and Cytosine Duplex DNA with Uninterrupted Spines of $Ag^{I}$-Mediated Base Pairs

Steven M. Swasey[1§], Frédéric Rosu[2§], Stacy M. Copp[3], Valérie Gabelica[4*] and Elisabeth G. Gwinn[5*]

[1] Department of Chemistry and Biochemistry, UCSB, Santa Barbara, CA 93117, USA

[2] Institut Européen de Chimie et Biologie, Université de Bordeaux, CNRS & Inserm (IECB, UMS3033, US001), 2 rue Robert Escarpit, 33607 Pessac, France.

[3] Center for Integrated Nanotechnologies, Los Alamos National Laboratories, Los Alamos, NM 87545, USA.

[4] Laboratoire Acides Nucléiques: Régulations Naturelle et Artificielle, Université de Bordeaux, Inserm & CNRS (ARNA, U1212, UMR5320), IECB, 2 rue Robert Escarpit, 33607 Pessac, France.

[5] Department of Physics, UCSB, Santa Barbara, CA 93117, USA

[§] Contributed equally to this work

* Correspondence to: Elisabeth G. Gwinn (bgwinn@physics.ucsb.edu) or Valérie Gabelica (v.gabelica@iecb.u-bordeaux.fr).

## Table of Contents

## Section S1. Full scan electrospray mass spectra

### S1.1. Methods

Experiments were performed on an Agilent 6560 DTIMS-Q-TOF instrument (Agilent Technologies, Santa Clara, CA), with the dual-ESI source operated in the negative ion mode. A syringe pump flow rate of 190 µL/h was used. Capacitance diaphragm gauges are connected to the funnel vacuum chamber and to the drift tube, and an additional flow controller admits Helium gas in the trapping funnel region. The flow controller is regulated by a feedback reading of the pressure in the rear of the drift tube. An in-house modification to the pumping system allows better equilibration of the pressures: a dry-compression multi-stage Roots vacuum pump Leybold ECODRY 40 plus (Leybold France SAS, Les Ulis, France) is connected to the source region with an Edwards SP16K diaphragm valve connected to the front pumping line, while the original Tri-scroll 800 pump is connected to the Q-TOF region. The helium pressure in the drift tube was 3.89 ± 0.01 Torr, and the pressure in the trapping funnel is 3.63 ± 0.01 Torr. The pressure differential between the drift tube and the trapping funnel ensures only helium is present in the drift tube.

The acquisition software version was B.07.00 build 7.00.7008.

All spectra were recorded using soft source conditions. The tuning parameters of the instrument (electrospray source, trapping region and post-IMS region (QTOF region) are optimized as described elsewhere.[1] The source temperature was set at 200°C and the source fragmentor voltage was set to 350 V. A fragmentor voltage of 600 V was also used only to allow better desolvation of the largest complexes, as mentioned in the figure legends. In the trapping funnel, the trap entrance grid delta was set to 2 V. The trapping time was 1000 µs and release time 200 µs. Trap entrance grid delta was set to 2V.

## S1.2. Supplementary figures

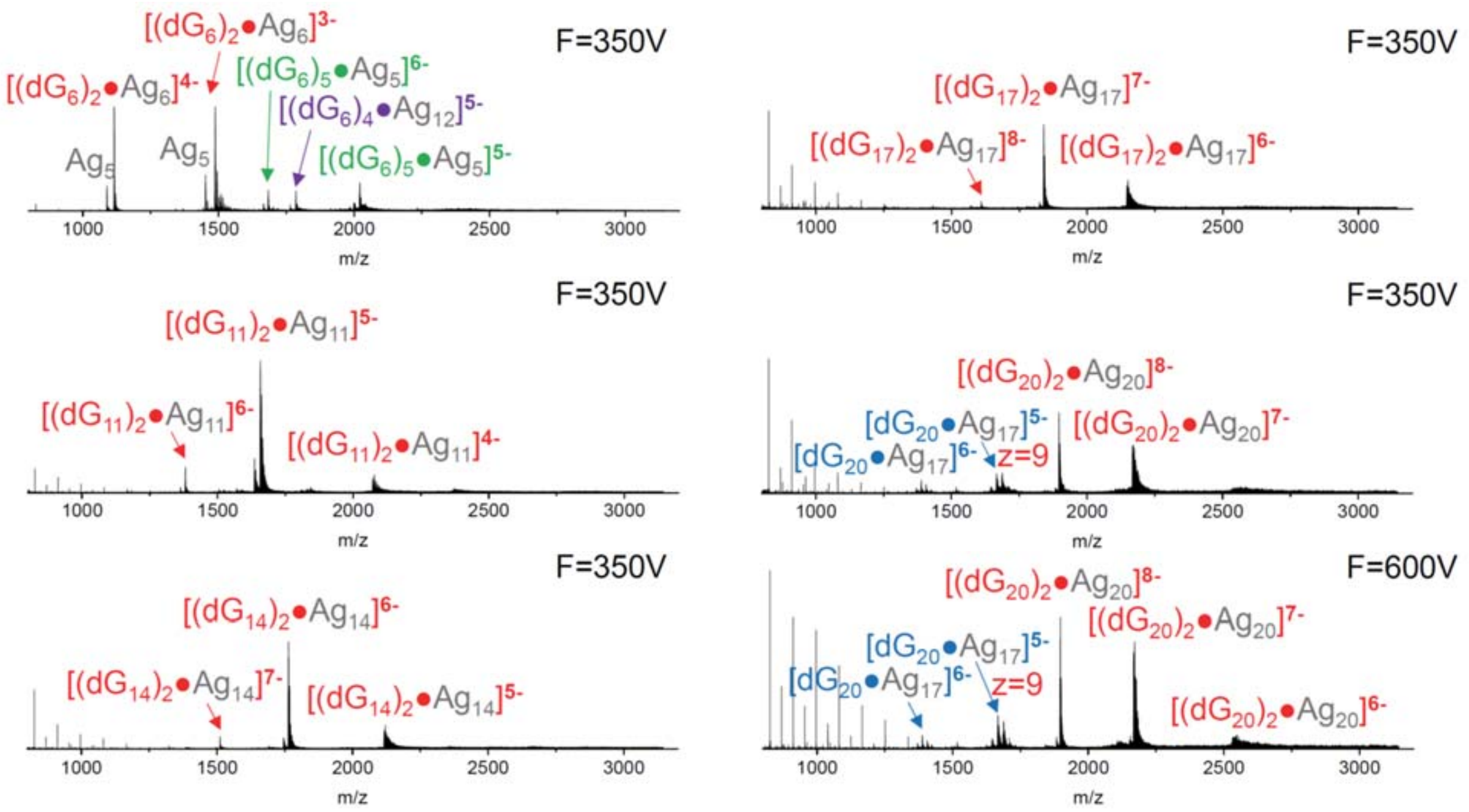

**Figure S1:** Full scan ESI-MS spectra of poly(dG) single strands prepared by annealing at 80 µM strand concentration with 1 equivalent of $AgNO_3$ per guanine, in 50 mM $NH_4OAc$. For the shortest strand ($dG_6$) in addition to the major product G-duplex in red, we found as minor products tetrameric structures (in purple, with 1 $Ag^I$ per 2 guanines), and pentameric structures (in green) having exactly five $Ag^I$ per pentamer, which could correspond to pentaplexes (G-quintets) with one $Ag^I$ ion in-between each quintet. Longer sequences (dG11 to dG20) did not form detectable higher-order structures. However, $dG_{20}$ starts to form intramolecular structures with more than one $Ag^I$ per pair of guanines (the main peak corresponds to 17 $Ag^I$ ions bound per strand). Because (i) these minor product peaks are already present in soft source conditions (fragmentor voltage = 350V) and (ii) the stoichiometry is much more than 10 $Ag^I$ ions, we believe they cannot come from gas-phase dissociation of the G-duplexes and must be formed also in solution. The equally spaced peaks at lower m/z correspond to $Ag_n(NO_3)_{n+1}$ clusters. The CCS distributions shown in Figure 3 of the main text were all recorded at a fragmentor voltage = 350V, and reconstructed on the peaks corresponding to the $[dG_n]_2{\bullet}Ag_n$ duplexes without extra $Ag^+$ or $NH_4^+$ adduct.

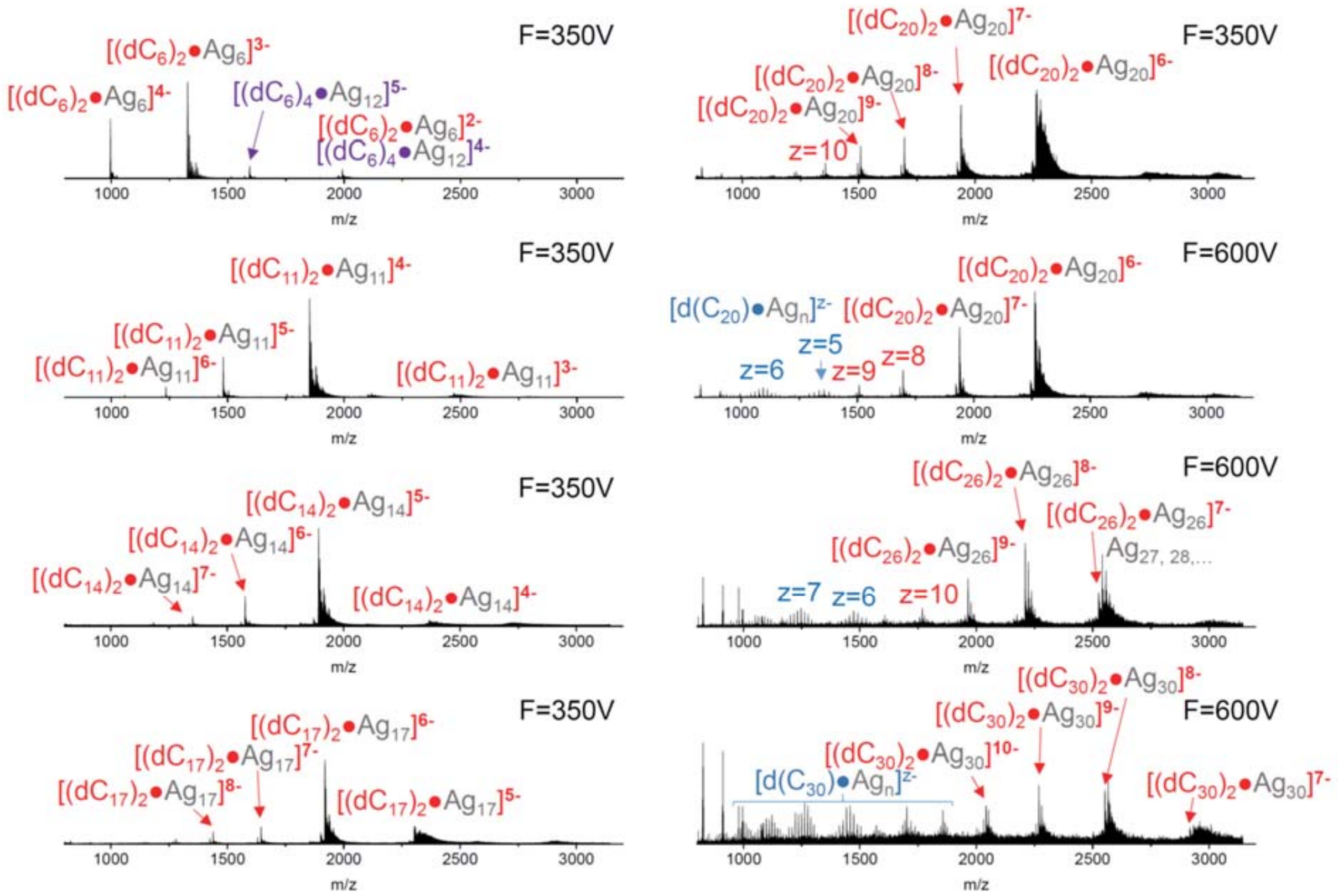

**Figure S2:** Full scan ESI-MS spectra of poly(dC) single strands prepared by annealing at 80 µM strand concentration with one equivalent of $AgNO_3$ per pair of cytosines in 50 mM $NH_4OAc$. The shortest strand ($dC_6$) formed tetramolecular structures in addition to the major product bimolecular structures. Up to $dC_{20}$, the spectra at fragmentor voltage = 350V were used to reconstruct the CCS distributions of the $[(dC_n)_2.Ag_n]$ complexes. For $dC_{26}$ and $dC_{30}$, the spectra were recorded at fragmentor voltage = 600V to obtain sufficient signal. For $dC_{20}$, signals corresponding to single strands with a distribution of $Ag^I$ adducts are detected at F=600V, not F=350V, suggesting that these single strands signals can come from partial gas-phase dissociation of the C-duplexes of the highest charge states in the harsher source conditions.

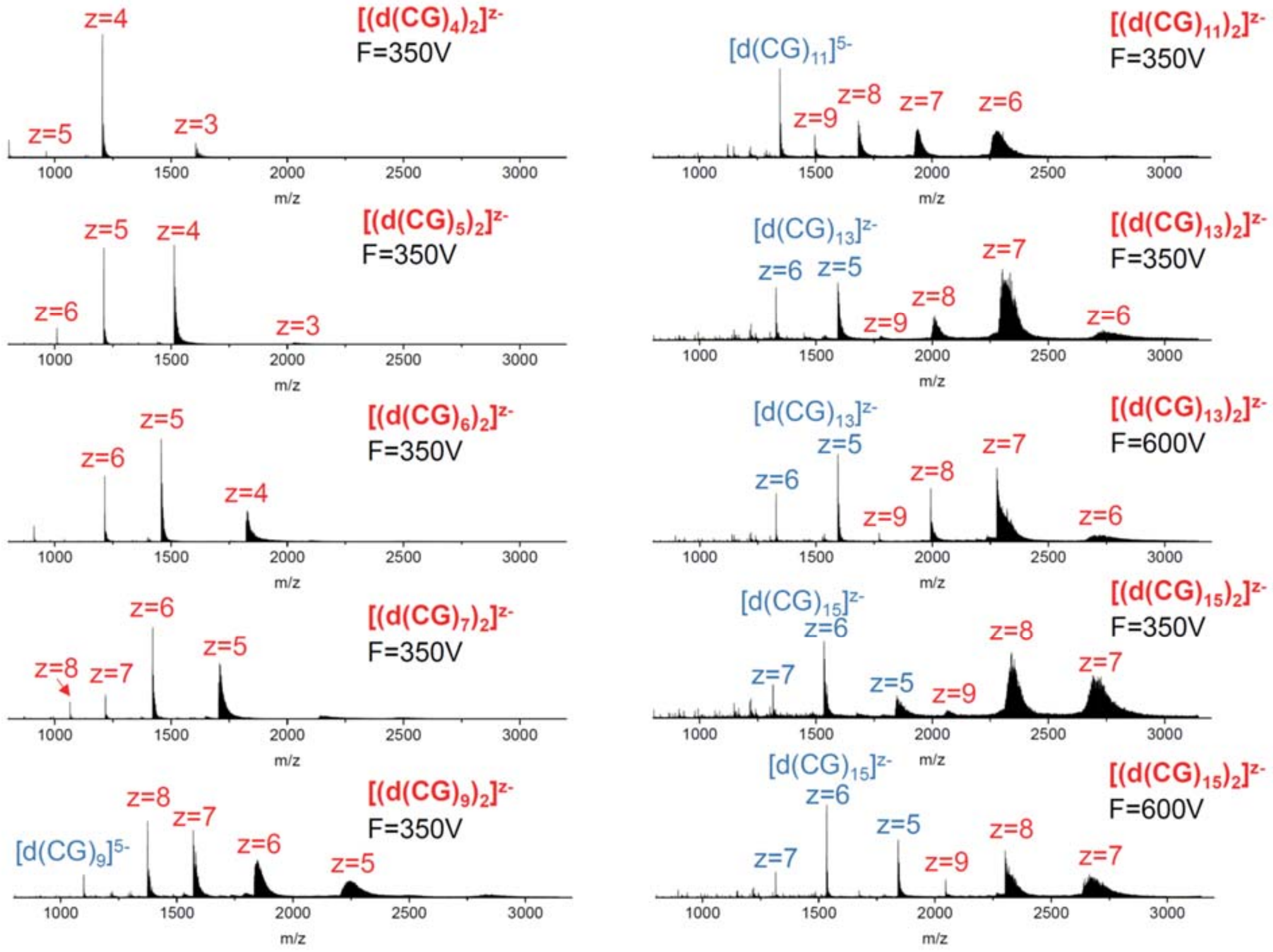

**Figure S3:** ESI-MS full scan spectra of poly(CG) duplexes, prepared at 20 µM duplex in 100 mM $NH_4OAc$. All spectra were recorded in soft source conditions, i.e. fragmentor voltage = 350V (F=350V). In addition, the two longest duplexes were also acquired at F=600V, because the nonspecific ammonium adducts were too numerous at 350V. When the peak with zero ammonium adducts was the dominant peak, the CCS distributions were reconstructed solely on that bare duplex. When the ammonium adduct distribution had a hump shape, the m/z range corresponding to the rising part on the left of the hump was integrated to reconstruct the CCS distributions shown in the main text. The CCS values of the peak centers were not significantly affected by the choice of number of adducts summed, but the signal-to-noise ratio was better for broad peaks when more adducts were summed. In Figure 3 of the main text, the F=350V data were used to reconstruct the CCS distributions of the 8-bp to 11-bp duplexes, and the F=600V data were used to reconstruct the CCS distributions of the 26-bp and 30bp duplexes.

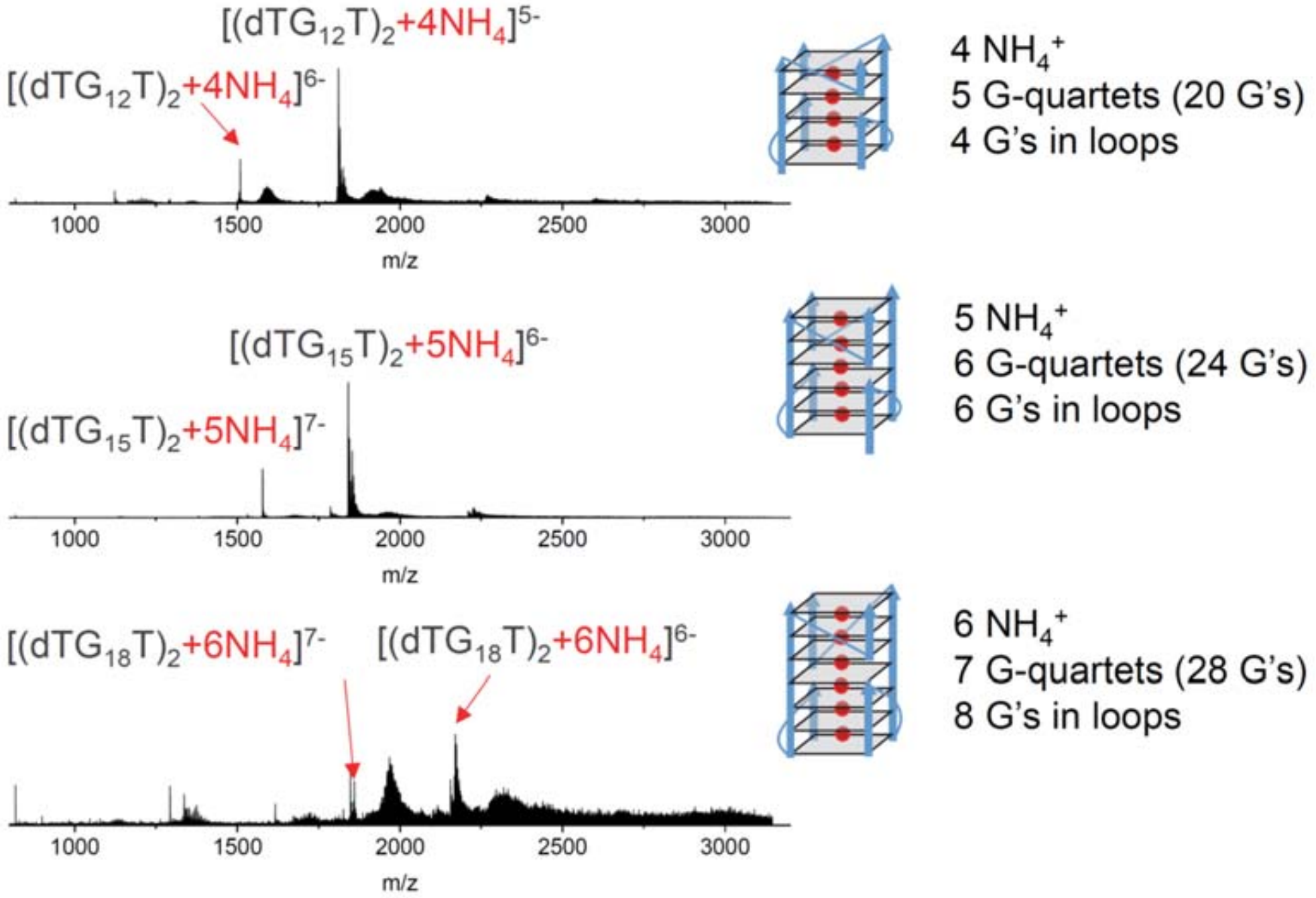

**Figure S4:** Full scan ESI-MS spectra of the bi-molecular G-quadruplexes prepared by annealing the single strands $dTG_nT$ at 100 µM in 150 mM $NH_4OAc$. The spectra of $TG_{12}T$ and $TG_{15}T$ were recorded at a fragmentor voltage of 350 V, and the spectrum of $TG_{18}T$ was recorded at a fragmentor voltage of 600 V. The cartoons show the number of G-quartets deduced from the number of ammonium cations specifically trapped in the structure.

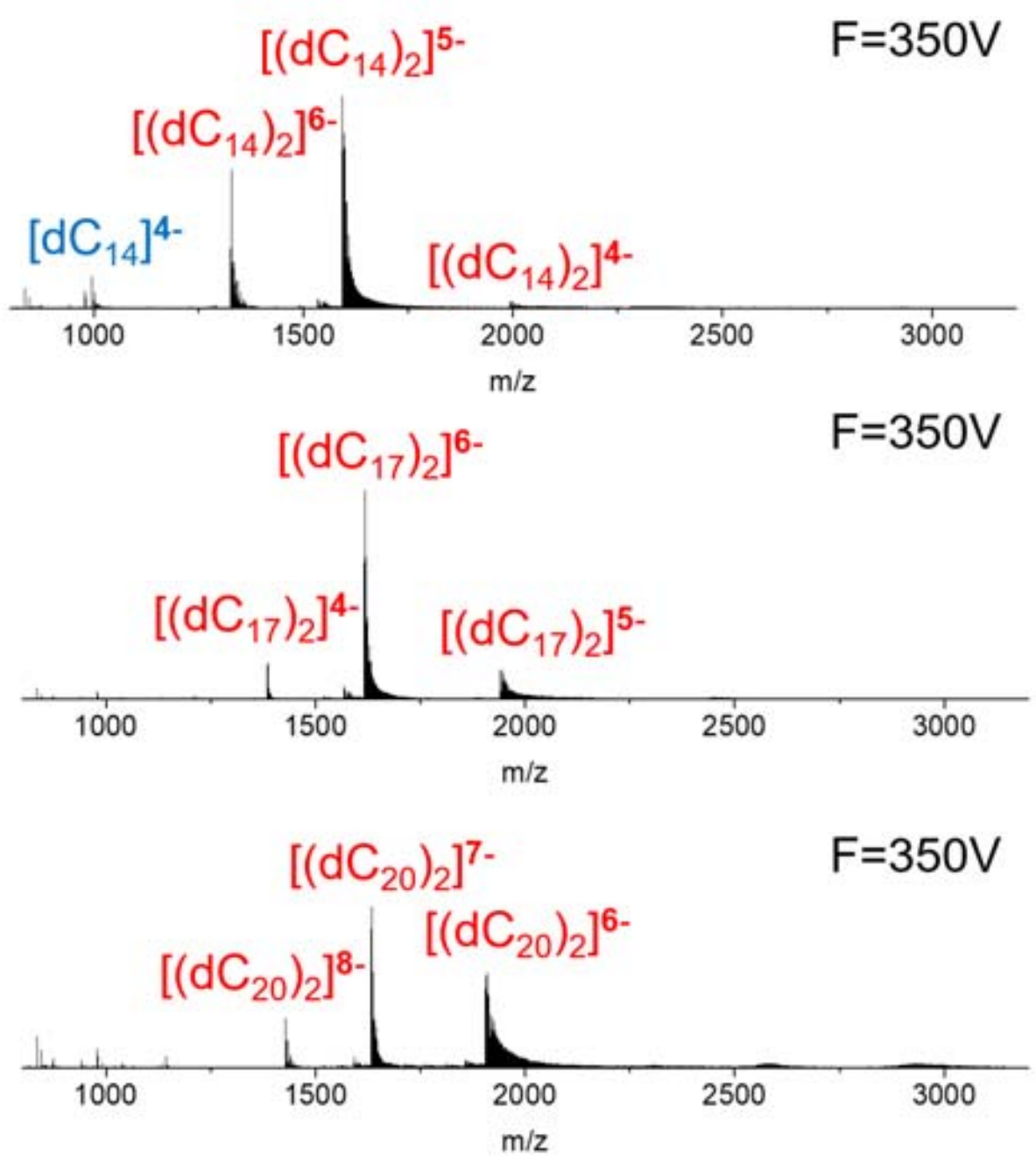

**Figure S5:** Full scan ESI-MS spectra of the bimolecular structures formed by $dC_n$ strands (50 µM) in presence of 100 mM $NH_4OAc$, without $AgNO_3$. At pH = 6.8, these structures are expected to form bimolecular i-motif structures. All spectra were recorded in soft conditions (fragmentor voltage = 350V) and the CCS distributions were extracted for the peaks without any nonspecific ammonium adduct.

**Section S2. FRET experiments**

**S2.1. Experimental Design**

$Ag^+$-paired strands were formed from strands labeled with Alexa 488 donors (″D″) and Alexa 647 acceptors (″A″) purchased with HPLC purification from Chemgenes. The dyes are attached to the DNA by $C_6$ linkers.

For the *cytosine* strand experiments, strand $\mathbf{A_{3'}}$ is 5'-$T_5C_{20}T_5$-3'-A (9814.0 g/mol), strand $\mathbf{D_{5'}}$ is D-5'-$T_2C_{20}T_2$-3' (7633.3 g/mol) and strand $\mathbf{D_{3'}}$ is 5'-$T_2C_{20}T_2$-3'-d (7665.3 g/mol).

For the *guanine* strand experiments, strand $\mathbf{A_{3'}}$ is 5'-$T_4G_{15}T_4$-3'-A (8360.5 g/), strand $\mathbf{D_{3'}}$ is 5'-$T_2G_{15}T_2$-3'-D (6818.8 g/mol) and strand $\mathbf{D_{5'}}$ is D-5'-$T_2G_{15}T_2$-3' (6818.7 g/mol). All masses were the reported values from the supplier Chemgenes, as confirmed by ESI-MS.

The donor emission and acceptor absorbance spectra correspond to a $R_0$ = 5.2 nm Förster radius, assuming randomized dye orientations. The dye spectra used to calculate $R_0$ are plotted in Figure S6. We measured these spectra in 10 mM $NH_4OAc$ pH = 7 for cytosine strands $\mathbf{A_{3'}}$ at 1.7 µM and $\mathbf{D_{5'}}$ at 1.6 µM.

To calculate $R_0$ we used the equation:[2,3]

$$R_0^6 = \frac{9\ln(10)\phi_D\kappa^2\left(\int f_D(\lambda)\varepsilon_A(\lambda)\lambda^4 d\lambda\right)}{128\pi^5 n^4 N_A} \quad (S1)$$

Here $\phi_D$ is the donor dye's fluorescence quantum yield (0.92 for D), $\kappa^2$ is the relative dipole orientation factor between the donor and acceptor dye (here we assume 0.66, corresponding to random orientation), $n$ is the medium's refractive index (1.33 for water), $N_A$ is Avagadro's number, and $\int f_D(\lambda)\varepsilon_A(\lambda)\lambda^4 d\lambda$ is the spectral overlap integral. In this integral, $f_D$ is the donor's emission spectrum normalized to an area of 1, and $\varepsilon_A$ is the acceptor's molar extinction coefficient spectrum.

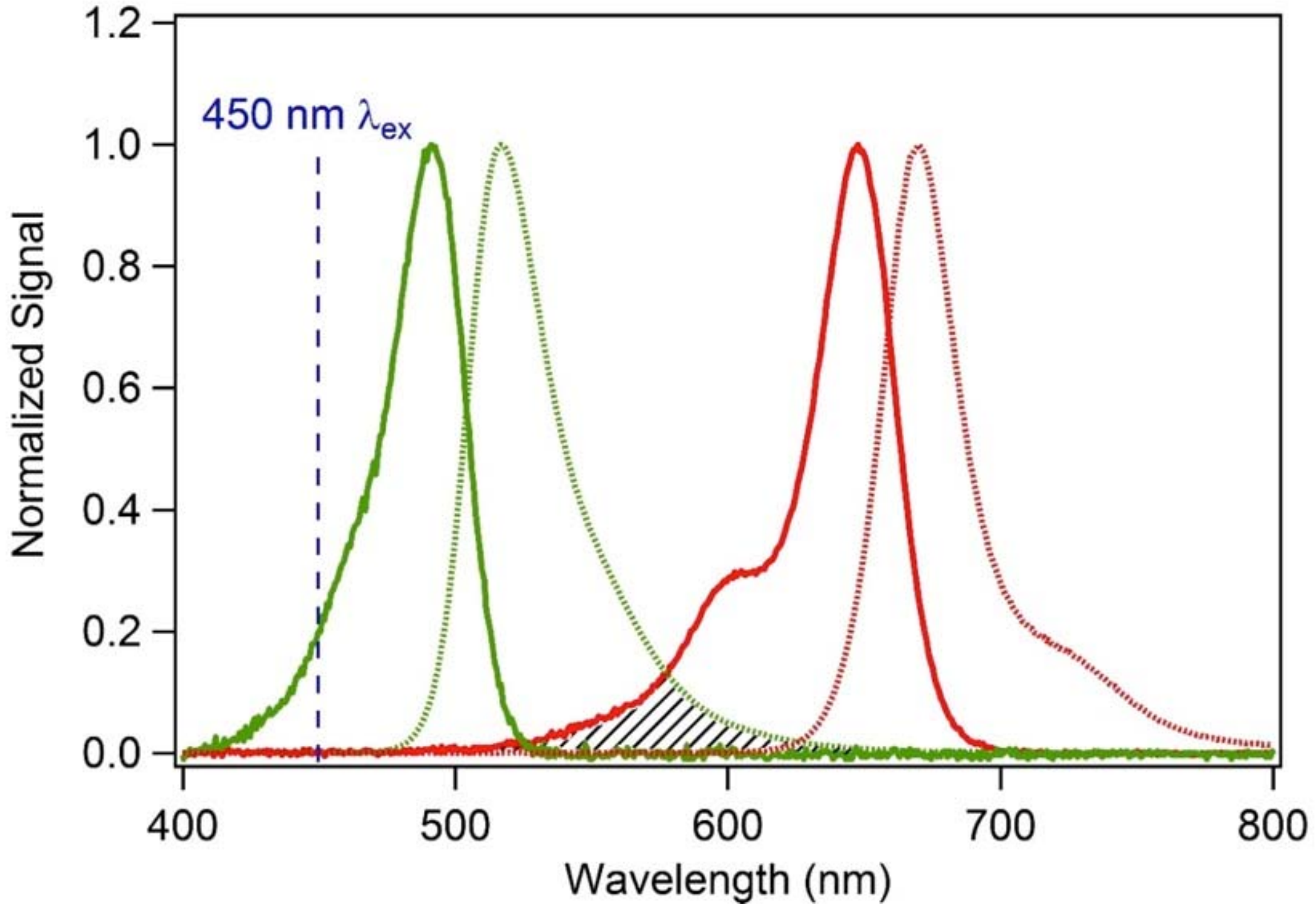

**Figure S6:** Experimental absorbance (solid curves) and emission (dashed curves) spectra for donor dye Alexa 488 “D” on strand D-5'-$T_2C_{20}T_2$-3' (green curves) and acceptor dye Alexa 647 “A” (red curves) on strand 5'-$T_5C_{20}T_5$-3'-A. Emission was excited at the visible absorbance peak in each case. The 450 nm excitation wavelength used for all FRET experiments (blue dashed line) negligibly excites the acceptor. The spectral overlap of the normalized emission spectrum of the donor and the normalized absorbance spectrum of the acceptor is represented by solid black slashed lines.

**S2.2. Purification and mass spectral characterization for the FRET experiments, and FRET experimental methods**

The dye-labelled DNA strands were synthesized and underwent one round of HPLC purification by ChemGenes Corporation. RNAse/DNAse free water (Integrated DNA Technologies) or HPLC/MS grade water was used in all of the experiments. The DNA strands were additionally desalted upon receipt by centrifugal solvent exchange with 3k MWCO filters.

To separate DNA products we used a Hitachi L-6200A pump and L-4200 UV-Vis detector with a 50 x 4.6 mm Kinetex EVO C18 column (2.6 μm particle size, 100 Å pore size) and a methanol and water gradient in 35 mM triethylamine ammonium acetate (pH 7). The DNA solutions were concentrated 3-fold prior to injection into the HPLC using 3k MWCO centrifugal filters. Both absorbance and emission (excitation at 270 nm) were monitored to determine which dyes were on each DNA product eluting from the column. After catching the separated HPLC peaks corresponding to $Ag^+$-paired strands with both acceptor and donor dyes, the solutions were solvent-exchanged by centrifugal filtration into 10 mM $NH_4OAc$ (for cytosine $Ag^+$-paired strands) or 50 mM $NH_4OAc$ (for guanine $Ag^+$-paired strands ) using 3k MWCO centrifugal filters. We used a 15-35% (cytosine $Ag^+$-paired strands) and 20-40% (guanine $Ag^+$-paired strands) methanol gradient at 1% per minute with a 1 μL/min flow rate for HPLC purification.

Spectral measurements at ambient temperature were performed for FRET studies. Samples were excited by Micropack's high power Xenon light source (HPX-2000) at 450 nm wavelength selected by a monochromator (Monoscan 2000). Emission was detected by a thermo-electrically cooled QE6500 Ocean Optics detector. The purified cytosine $Ag^+$-paired strand solutions were injected into a Waters QTOF2 negative-ion mode electrospray-ionization (ESI) mass-spectrometer at 10 μL/min with a 2 kV capillary voltage, 30V cone voltage and 10V collision energy for product verification. The purified guanine $Ag^+$-paired strand solutions were injected into a Xevo G2-XS QToF ESI mass spectrometer in negative-ion mode at 5 μL/min with a 2kV capillary voltage, 10V cone voltage and 6V collision energy for product verification.

### S2.2.1. Cytosine DNA strands

The cytosine $Ag^+$-paired strands were synthesized by separately combining strands **A3´** (5´-d($T_5C_{20}T_5$)-[Alexa 647]-3´) with **D5´** (5´-[Alexa 488]-d($T_2C_{20}T_2$)-3´), and **A3´** with **D3´** (5´-d($T_2C_{20}T_2$)-[Alexa 488]-3´), at a final concentration of 2.5 μM for each DNA strand in a solution of 500 mM $NH_4OAc$ (pH 7) and 100 μM $AgNO_3$ (Sigma-Aldrich analytical grade). The solutions were annealed at 90 °C for 5 minutes, and allowed to cool slowly to room temperature. After cooling, the samples were purified by HPLC. FRET and ESI-MS measurements were performed on the purified **D3´•A3´** and **D5´•A3´** duplexes as detailed by the description and data below.

Here we use C(**3**´-3´) to denote the unpurified solution produced by mixing cytosine strands **D3´** and **A3´** with $Ag^+$, and C(**5**´-3´) to denote the unpurified solution produced by mixing cytosine strands **D5´** and **A3´** with $Ag^+$.

We performed HPLC on the C(**3**´-3´) and C(**5**´-3´) solutions to isolate the $Ag^+$-paired **D3´•A3´** and **D5´•A3´** products from the mixture of A-A, D-D and D-A products formed by $Ag^+$-mediated assembly. Figure S7 shows the full HPLC absorbance and emission chromatograms. We identified the HPLC peaks corresponding to **D•A** products using UV excitation and monitoring the emission chromatograms for simultaneous emission from D and A (labelled D-A peaks, Fig. S7b,d). Figure S8 shows the corresponding mass spectra for the purified **D•A** products collected during elution of the peaks labelled D-A in the HPLC emission chromatograms in Figs. S7b,d.

The data for C(**3**´-3´) exhibits one HPLC peak with time correlated emission from both D and A (labelled $D_{3´}$-$A_{3´}$ in Fig. S7b), which was captured for FRET measurements (main text Fig. 1b, solid blue curve).

The bumpy shape of this peak indicates that it contains silver-paired **D3´•A3´** with slightly different conformations. The dominant product detected in ESI-MS of this eluent peak contains one **D3´** strand, one **A3´** strand and 19 $Ag^+$ (Figure S8a) and the neighboring, lower abundance peak is **D3´•A3´** with 18 $Ag^+$. In addition, the ESI-MS for **D3´•A3´** contains still lower abundance peaks, indicated by "*" in Fig. S8a, that have the same silver content but are missing one T base from the **D3´** strand. This missing T base likely contributes to the peak substructure of the **D3´•A3´** HPLC peak (Figure S7a,b; structural perturbations due to donor dye interactions with the DNA may also contribute to the **D3´•A3´** HPLC peak substructure). The presence of **D3´** strands containing such single T base deletions should not significantly affect the measured FRET efficiencies due to the small change in separation of the donor and acceptor, as supported by the consistent ratio of acceptor to donor dye emission across the entire peak structure in Figure S7b.

The data for C(**5**´-3´) exhibits two closely spaced HPLC peaks with time-correlated emission from both D and A, labelled $D_{5'}$-$A_{3'}$ and $(D_{5'}\text{-}A_3)_y$ in Fig. S7d. For the $D_{5'}$-$A_{3'}$ HPLC peak collected and used for FRET experiments (main text Fig. 2a, orange curve), the dominant product detected in ESI-MS contains 18 $Ag^+$ (Figure S8b). For the $(D_{5'}\text{-}A_3)_y$ HPLC peak, the major product is also a **D5´•A3´** product with 18 $Ag^+$ (Figure S8c). ESI-MS of the $(D_{5'}\text{-}A_3)_y$ HPLC peak showed contamination by $Ag^+$-paired **D5´•D5´** which would contribute a spurious background signal in FRET from the donor dye excitation, so we did not perform FRET experiments on $(D_{5'}\text{-}A_3)_y$. We believe the two separate HPLC peaks labelled $D_{5'}$-$A_{3'}$ and $(D_{5'}\text{-}A_3)_y$ in Fig. S7d, likely arise from structural perturbations caused by the donor dye, possibly intercalation or some other mechanism.

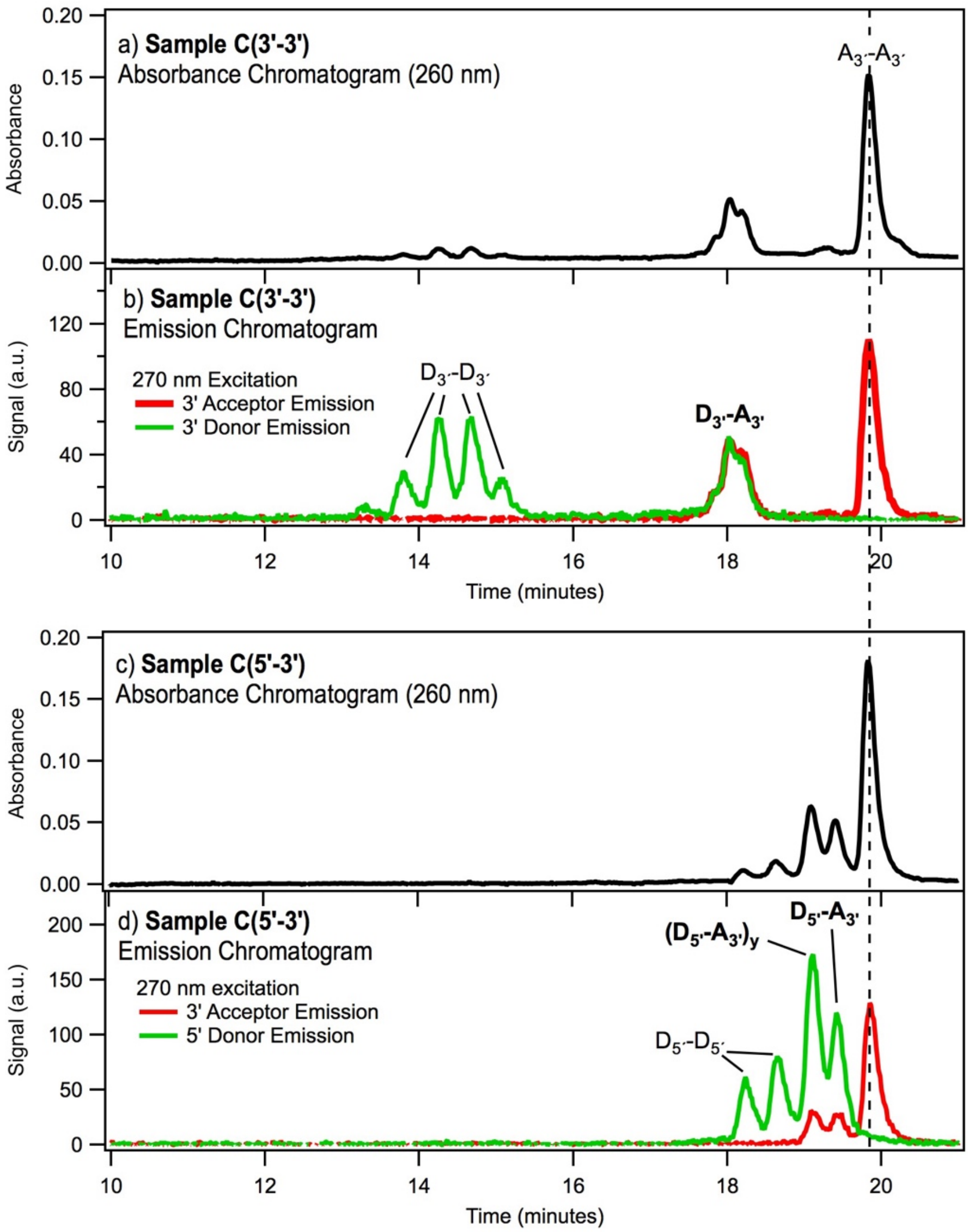

**Figure S7:** (a),(b) HPLC chromatograms for sample C(**3**′-3′), the unpurified solution containing strands $D_{3'}$ and $A_{3'}$ with $Ag^+$, for a) absorbance and b) emission. (c),(d) HPLC chromatograms for sample C(**5**′-3′), the unpurified solution containing strands $D_{5'}$ and $A_{3'}$ with $Ag^+$) for (c) absorbance and (d) emission. We used 270 nm excitation to simultaneously excite both the donor (green curves, (b) and (d)) and acceptor (red curves, (b) and (d)). Peaks with signal from both the donor and acceptor dyes in the emission chromatograms in (b) and (d) are the D-A products which contain both the acceptor and donor dye-labelled strands. The peaks marked by black, dashed lines are the $Ag^+$-paired $A_{3'}$-$A_{3'}$. Peaks for the $Ag^+$-paired $D_{3'}$-$D_{3'}$ (b) or $D_{5'}$-$D_{5'}$ (d) are also labeled. As shown in (b), there is only one HPLC peak for the $Ag^+$-paired $A_{3'}$-$A_{3'}$(red curve, peak eluting near 20 minutes) while the $Ag^+$-paired $D_{3'}$-$D_{3'}$ and $D_{5'}$-$D_{5'}$ each have multiple peaks associated with them. Apparently the donor dye produces distinct, minor structural perturbations in the $Ag^+$-paired D-D. Therefore similar structural perturbations can be expected for the $Ag^+$-paired D-A. The substructure in the $D_{3'}$ -$A_{3'}$ peak in b) also indicates the presence of minor structural variants, due to the presence of strands with a missing thymine nucleotide (see mass spectra in Figure S8a) as well as possible donor dye structural perturbations. However, the ratio of acceptor to donor dye emission remains constant across the entire $D_{3'}$-$A_{3'}$ peak, indicating that the structural variations are slight (otherwise FRET would have been altered), as expected given the closely spaced elution times. In (d), for solutions of $A_{3'}$ and $D_{5'}$ with $Ag^+$, there are two slightly separated peaks with concurrent donor and emission, likely also arising from structural perturbations from the donor dye. The slightly slower eluting peak on the right is the $Ag^+$-paired $D_{5'}$-$A_{3'}$ discussed in the main text (Figure 2a). ESI-MS of the peak labelled $(D_{5'}$-$A_{3'})_y$ showed contamination with co-eluting $Ag^+$-paired $D_{5'}$-$D_{5'}$, precluding FRET analysis.

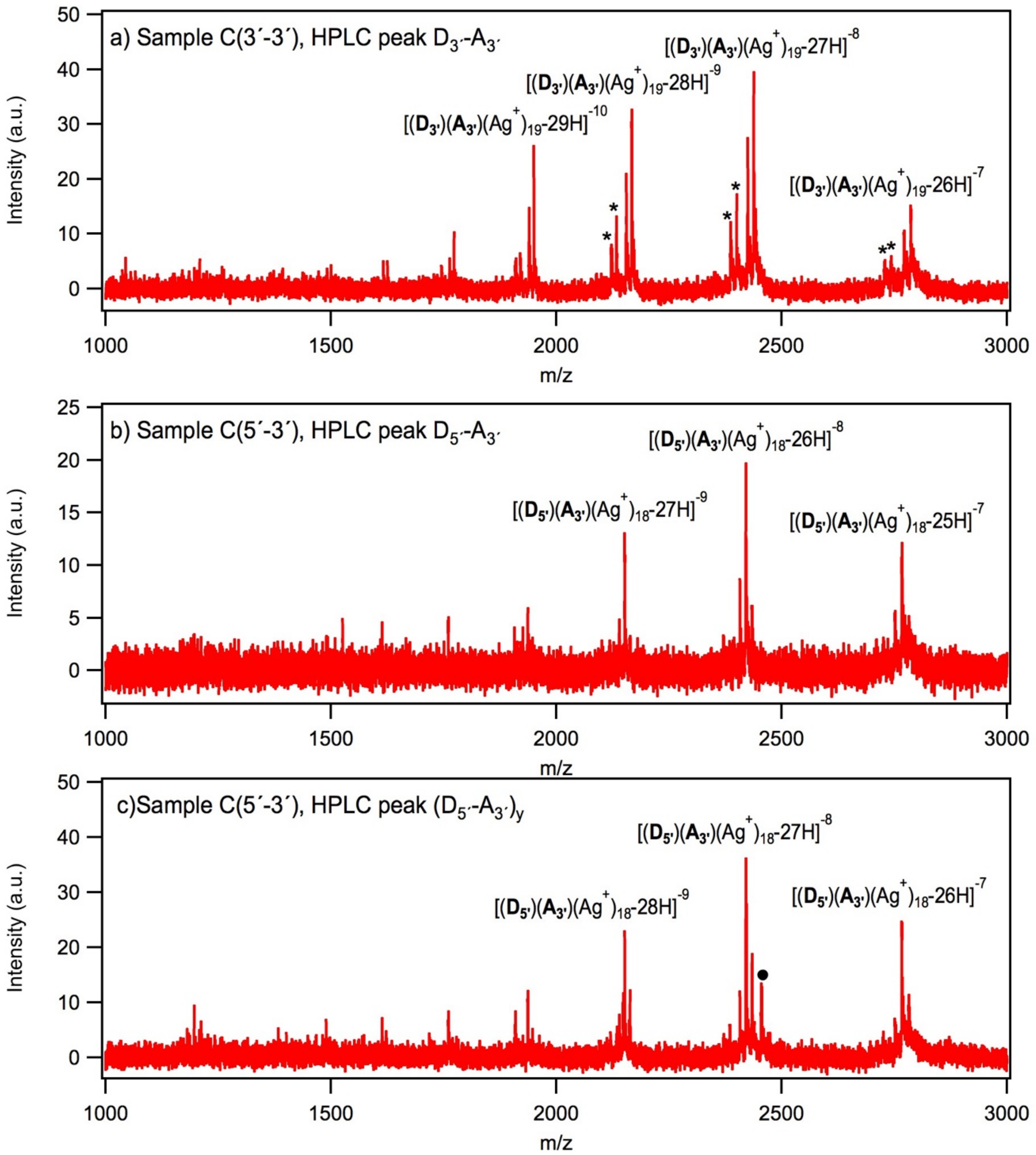

**Figure S8:** Mass spectra of the HPLC-isolated D-A product peaks marked in Fig. S7. Main products are labelled with the overall charge state of the ion, the number of $Ag^+$ and the number of protons removed.

(a) ESI-MS for the $D_{3'}$-$A_{3'}$ HPLC peak isolated from sample C(3´-3´). The dominant product is $D_{3'}$-$A_{3'}$ with 19 $Ag^+$ ([$D_{3'}$)($A_{3'}$)( $(Ag^+)_{19}$-26H]$^{-8}$ is at 2438.01 m/z). The lower abundance peak immediately to the left of the dominant product peak is $D_{3'}$-$A_{3'}$ with 18 $Ag^+$. The "*" symbols mark $D_{3'}$-$A_{3'}$ with 18 and 19 $Ag^+$ that are missing one thymine nucleotide.

(b) ESI-MS for the $D_{5'}$-$A_{3'}$ HPLC peak isolated from sample C(5´-3´). The dominant product is $D_{5'}$-$A_{3'}$ with 18 $Ag^+$ ([($D_{5'}$)($A_{3'}$)]$(Ag^+)_{18}$-27H]$^{-8}$ is at 2420.70 m/z).

(c) ESI-MS for the ($D_{5'}$-$A_{3'}$)$_y$ HPLC peak isolated from sample C(5´-3´). The dominant product is also $D_{5'}$-$A_{3'}$ with 18 $Ag^+$ ([($D_{5'}$)($A_{3'}$)]$(Ag^+)_{18}$-27H]$^{-8}$ is at 2420.72 m/z). We also find a minor peak from $Ag^+$-paired $D_{5'}$-$D_{5'}$ with 18 $Ag^+$ (2454.89 m/z; marked with a closed circle) which elutes just before (Fig.S7d). Also present as minor products are $D_{5'}$-$A_{3'}$ $Ag^+$-paired strands with 18 $Ag^+$.

In (a)-(c), no other products are detected at significant abundance for $A_{3'}$-$D_{3'}$ or $D_{5'}$-$A_{3'}$. Thus the number of strands without dye molecule labels appears to be negligible in the purified aliquots used for FRET experiments.

### *S2.2.2. Guanine DNA strands*

The guanine **A3´** strand is 5´-d($T_4G_{15}T_4$)-[Alexa 647]-3´, **D3´** is 5´-d($T_2G_{15}T_2$)-[Alexa 488]-3´ and **D5´** is 5´-[Alexa 488]-d($T_2G_{15}T_2$)-3´.

The preparation steps described above for the cytosine case produced insufficient yields of $Ag^+$-paired guanine **D•A** duplexes for FRET experiments, requiring us to develop alternate procedures. In the guanine case, the additional challenges included the polydispersity of the strands received from the manufacturer, arising from the difficulty of synthesizing DNA with long runs of G bases; aggregation of the strands; and the pronounced tendency of the strands to self-pair upon addition of $Ag^+$ (ie, form **D•D** and **A•A** duplexes rather than **D•A** duplexes).

The initial step in overcoming these challenges was to break up the $\mathbf{A_{3'}}$ aggregates in the solution provided by the manufacturer (Fig. S9a). We achieved this by adding $Ag^+$ (375 μM ) to the 25 μM $\mathbf{A_{3'}}$ DNA in $NH_4OAc$ (pH 7) to form high yields of unaggregated, $Ag^+$-paired **A•A** duplexes(Fig. S9b). ESI-MS (not shown) revealed the presence of truncated (shorter) strands. To isolate the desired, pure **A3'** strand, we injected the polydisperse $Ag^+$-paired **A3′** strands into the HPLC and collected aliquots of all resolvable major peaks. The products were examined by ESI-MS to determine which HPLC peak contained the $Ag^+$-paired **A•A** duplex of the full length strand, 5′-d($T_4G_{15}T_4$)-[Alexa 647]-3′(Figure S11a). This purified, full length **A•A** aliquot was then incubated in 50 mM cysteine for 1 hour to chelate out the $Ag^+$, solvent exchange to re-establish buffer conditions, and then used in the subsequent steps, described below, for the preparation of $Ag^+$-paired guanine **D•A** duplexes. The procedure of $Ag^+$-pairing followed by chelation was also used to disaggregate the **D3′** and **D5′** guanine strands.

Initial attempts to create $Ag^+$-paired **D•A** duplexes by mixing these purified **A** and **D** solutions failed: after addition of $AgNO_3$, $Ag^+$-paired **A•A** and **D•D** products overwhelmingly predominated, presumably due to strand self-pairings in the individual **A** and **D** solutions. To obtain sufficient yields of the **D•A** pairs for FRET studies, we synthesized $Ag^+$-paired **D•D** and **A•A** using 75 μM $AgNO_3$, 2.5 μM **D3′** or **D5′**, and 2.5 μM **A3′** (disaggregated and purified as described above) in 50 mM $NH_4OAc$ (pH 7), and annealed them at 90 °C for 5 minutes, then cooled slowly to room temperature. After cooling, aqueous cysteine was added to each solution to a final concentration of 50 mM, followed by solvent exchange to 50 mM $NH_4OAc$ using centrifugal filtration with 3k MWCO filters to remove excess cysteine (approximately 0.001% of original solvent remained). These disaggregated **D** and **A** strands were mixed and brought to a final concentration of 75 μM $AgNO_3$, then annealed again.

In the remainder of this section we use G(**3′**-3′) to denote the unpurified solution produced by mixing the purified and disaggregated guanine strands **D3′** and **A3′** with $Ag^+$, and G(**5′**-3′) to denote the unpurified solution produced by mixing the purified and disaggregated guanine strands **D5′** and **A3′** with $Ag^+$. We performed HPLC on the G(**3′**-3′) and G(**5′**-3′) solutions to isolate the $Ag^+$-paired **D3′•A3′** and **D5′•A3′** products from the mixture of **A•A**, **D•D** and **D•A** products formed by $Ag^+$-mediated assembly. Figure S10 shows the full HPLC absorbance and emission chromatograms. We identified the HPLC peaks corresponding to **D•A** products using UV excitation and monitoring the emission chromatograms for correlated emission from D and A (labelled "D-A" peaks, Fig. S10b,d). Figure S11b-d shows the corresponding mass spectra for the purified **D•A** products collected during elution of the D-A labelled peaks in the HPLC emission chromatograms in Fig. S10b,d. (The **D•A** aliquots collected during HPLC were solvent exchanged by spin centrifugation (3k MWCO filters) into 50 mM $NH_4OAc$ and 10 μM $AgNO_3$. The guanine $Ag^+$-paired **D•A** duplexes required this slight addition of $AgNO_3$ to retain high yields of duplexes in ESI-MS, as opposed to the cytosine $Ag^+$-paired **D•A** duplexes which were stable post-HPLC with no additional $AgNO_3$. In both guanine and cytosine cases, without HPLC the duplex products are stable). After solvent exchanging, the samples were immediately spectrally analyzed for FRET signals and then injected into the ESI-MS to verify product composition (Fig. S11b-d).

The data for G(3′-3′) exhibits one HPLC peak with time-correlated emission from both D and A (labelled $D_{3'}$-$A_{3'}$ in Fig. S10b), which was captured for FRET measurements (main text Fig. 2b, solid blue curve). The dominant product detected in ESI-MS of this eluent peak contains one **D3′** strand, one **A3′** strand and 15 $Ag^+$ (Figure S11b). The secondary peak, labelled with a "*" in Fig. S11b, is the same product missing a donor dye. This does not affect FRET analysis because **D•A** pairs without donor dyes have no emission (acceptor or donor) for the 450 nm excitation wavelength used in FRET measurements.

The data for G(**5**′-3′) exhibits two HPLC peaks with time-correlated emission from both D and A. The main HPLC peak, labelled $D_{5'}$-$A_{3'}$ in Figure S10d, was captured for FRET measurements (main text Fig. 2b, solid orange curve). The dominant product detected in ESI-MS of this eluent peak contains one **D5′** strand, one **A3′** strand and 15 $Ag^{+}$ (Figure S11c). For the lower abundance peak labelled labeled $(D_{5'}$-$A_{3'})_{y}$ in Figure S10d, ESI-MS (Figure S11d) also showed a dominant product with one **A3'** strand, one **D5'** strand and 15 $Ag^{+}$ as well as the same product with a donor dye missing. We believe that structural perturbations caused by the donor dye are responsible for the two separate HPLC peaks, $D_{5'}$-$A_{3'}$ and $(D_{5'}$-$A_{3})_{y}$ labelled in Fig. S10d.

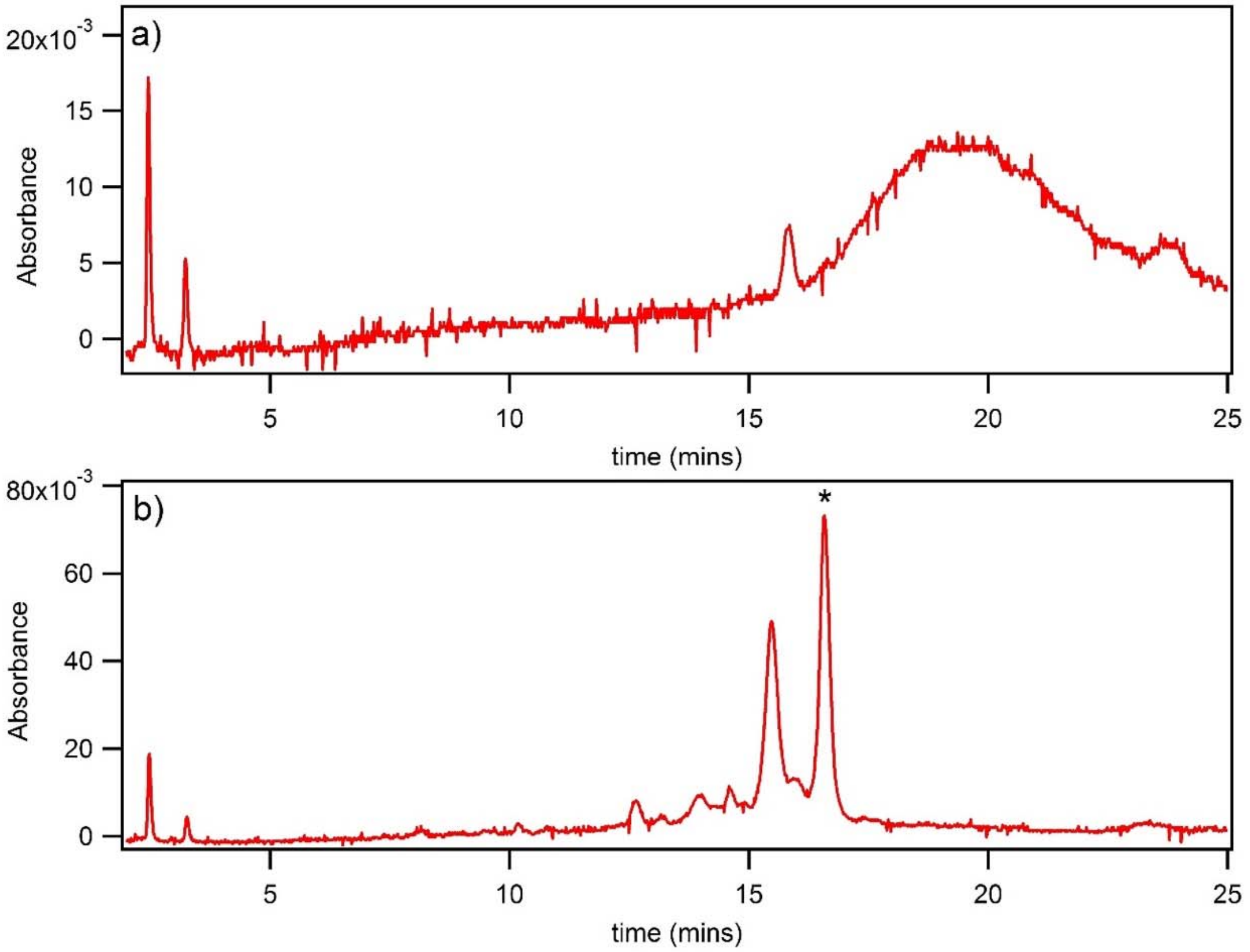

**Figure S9:** HPLC absorbance chromatograms monitored at 260 nm for guanine strand $A_{3'}$(d($T_4G_{15}T_4$)-[Alexa 647]-3´) at 2.5 µM with a) 10 mM $NH_4OAc$ and b) 50 mM $NH_4OAc$ and 37.5 µM $AgNO_3$. The broad and relatively featureless chromatogram in a) with no added $AgNO_3$ indicates aggregation. Conversely, in b) the addition of $AgNO_3$ produces well defined elution peaks. The peak denoted with a "*" was identified by ESI-MS (Figure S11a) as $Ag^+$-paired $A_{3'}$-$A_{3'}$ containing 15 $Ag^+$. Apparently, adding $AgNO_3$ breaks up the bare aggregated guanine strands to form unaggregated $Ag^+$-paired guanine duplexes.

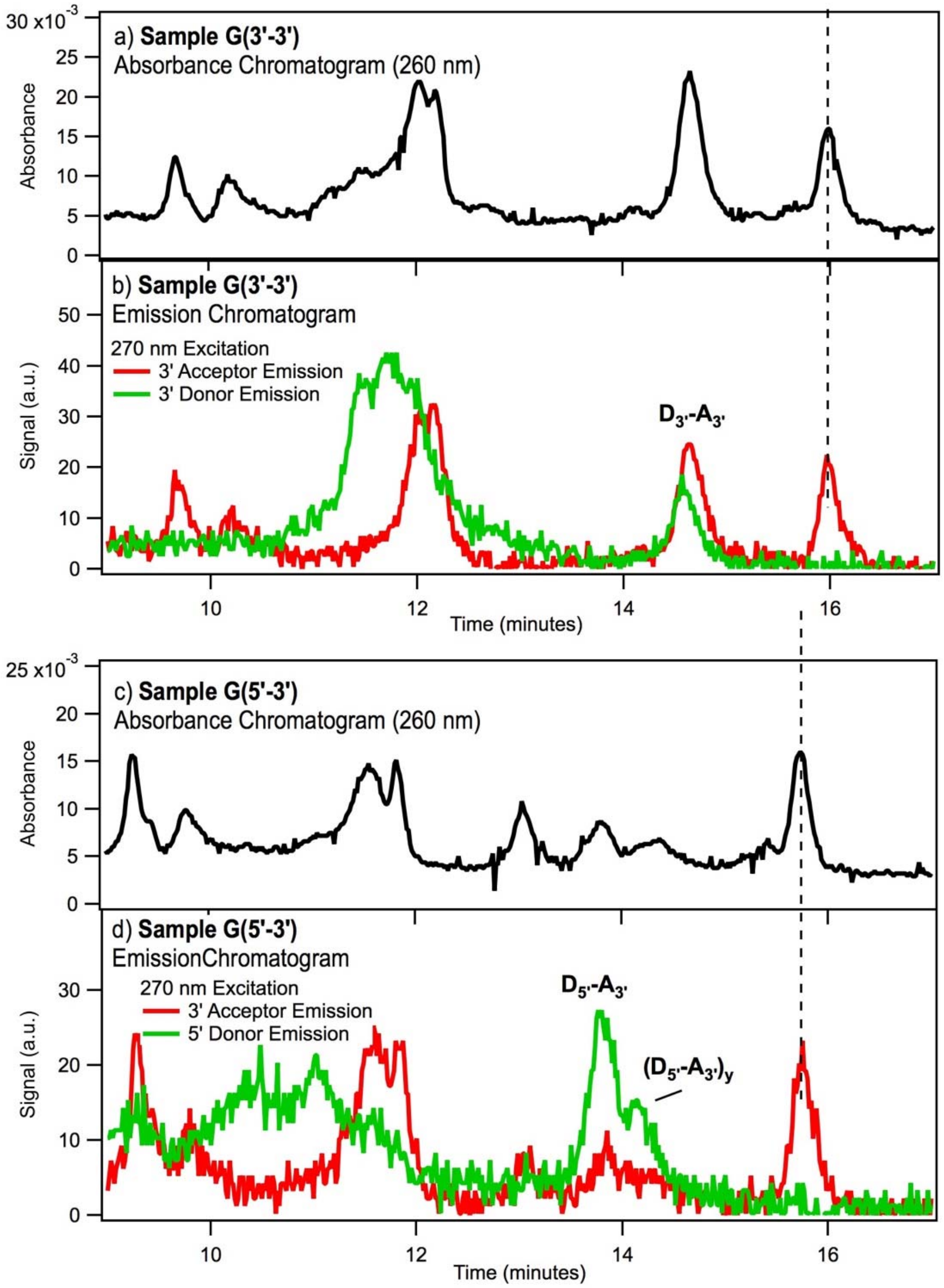

**Figure S10:** (a),(b) HPLC chromatograms for sample G(**3**'-3'), the unpurified solution containing strands $A_{3'}$ and $D_{3'}$ with $Ag^+$, for a) absorbance and b) emission. (c),(d) HPLC chromatograms for sample G(**5**'-3'), the unpurified solution containing strands $A_{3'}$ and $D_{5'}$ with $Ag^+$, for c) absorbance and d) emission. We used 270 nm excitation to simultaneously excite both the donor dye (green curves, (b) and (d)) and acceptor dye (red curves, (b) and (d)). Black dashed lines mark elution of $Ag^+$-paired $A_{3'}$-$A_{3'}$. The chromatograms are more structured than for the cytosine case, reflecting greater product heterogeneity that results in multiple time intervals with both donor and acceptor emission due to partially overlapping elution of various $Ag^+$ paired $D_{3'}$-$D_{3'}$, $D_{5'}$-$D_{5'}$, or monomeric strand products with $Ag^+$. Intervals with different time signatures for acceptor and donor emission do not correspond to D-A pairs. The D-A labelled chromatogram peaks in (b) and (d) have correlated emission for both the donor and acceptor dyes and were verified by ESI-MS to be $Ag^+$-paired D-A products (Fig. S11b,d). The $D_{\mathbf{3}'}$-$A_{3'}$ peak in (b) and the $D_{\mathbf{5}'}$-$A_{3'}$ peak in (d) were captured during HPLC and used for FRET experiments (Fig. 2b, main text).

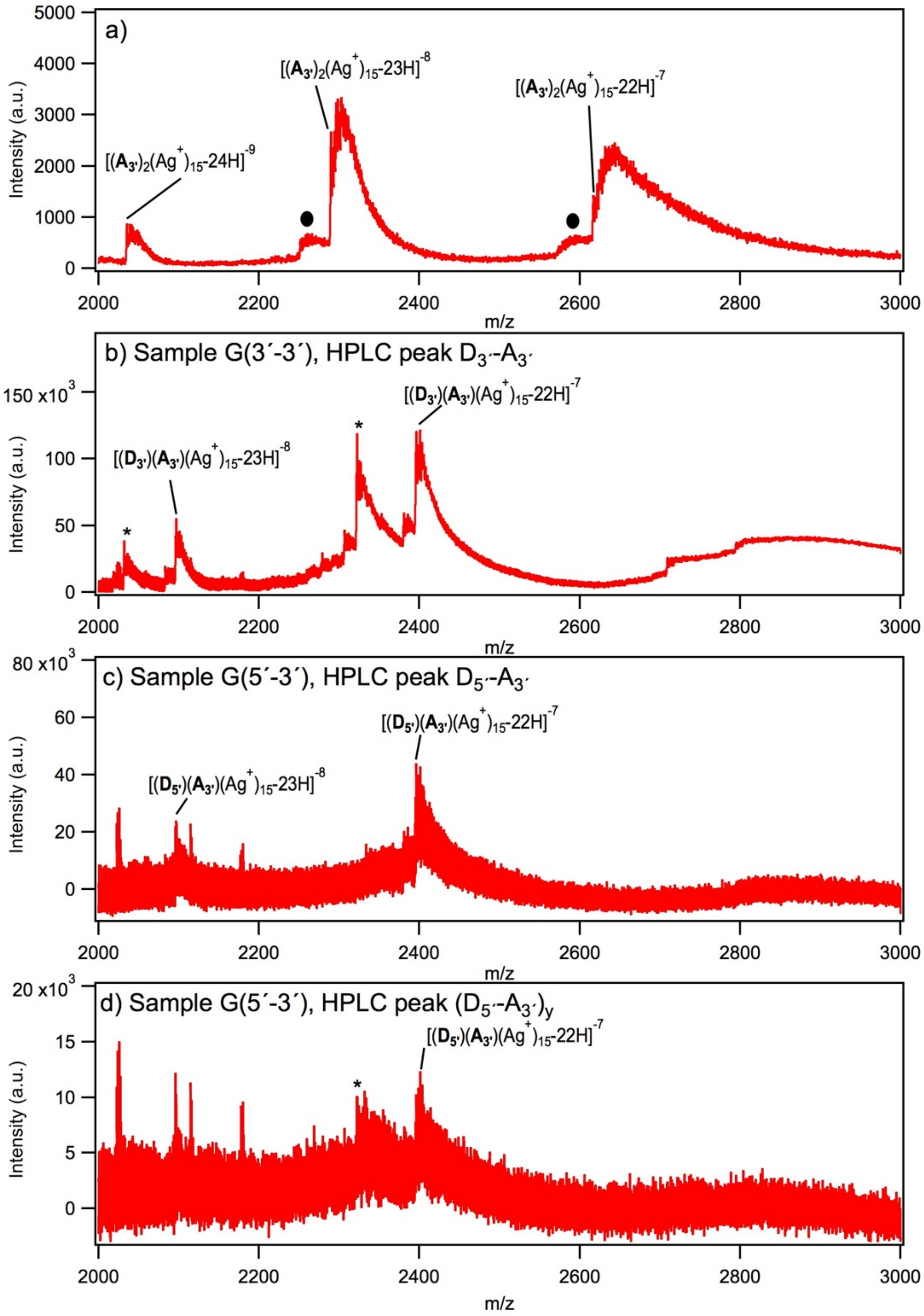

**Figure S11:** Mass spectra of HPLC purified solutions of the guanine $Ag^+$-paired products. Main products are labelled with the overall charge state of the ion, the number of $Ag^+$ and the number of protons removed. (a) Mass spectrum of the HPLC-isolated $A_{3'}$-$A_{3'}$ peak labeled with a "*" in Fig. S9b. The most abundant product in is $A_{3'}$-$A_{3'}$ with 15 $Ag^+$ ($[(\mathbf{A_{3'}})_2(Ag^+)_{15}\text{-}23H]^{-8}$ at 2289.80 m/z), with a broad tail of salt peaks at higher m/z. Impurity products marked by black circles are the same product except that one thymine nucleotide is missing. (b) Mass spectrum of the HPLC-isolated $D_{3'}$-$A_{3'}$ peak (Figure S10b) used in FRET experiments. The labeled major product is $A_{3'}$-$D_{3'}$ with 15 $Ag^+$ ($[(\mathbf{A_{3'}})(\mathbf{D_{3'}})(Ag^+)_{15}\text{-}22H]^{-7}$ at 2396.14 m/z), with trailing salt peaks at higher m/z. Peaks marked by "*"are the same product except missing the donor dye, which may have fragmented off during ESI. Because the 450 nm excitation wavelength used in the FRET experiments does not excite the acceptor dye, $Ag^+$-paired strands missing a donor dye do not affect the FRET analysis. (c) Mass spectrum of the HPLC-isolated $D_{5'}$-$A_{3'}$ peak (Figure S10d) used in FRET experiments. The dominant product is $D_{5'}$-$A_{3'}$ with 15 $Ag^+$ ($[(\mathbf{D_{5'}})(\mathbf{A_{3'}})(Ag^+)_{15}\text{-}22H]^{-7}$ at 2396.11 m/z). (d) Mass spectrum of the HPLC-isolated $(D_{5'}\text{-}A_{3'})_y$ peak (Fig. S10d). The dominant product is $D_{5'}$-$A_{3'}$ with 15 $Ag^+$ ($[(\mathbf{D_{5'}})(\mathbf{A_{3'}})(Ag^+)_{15}\text{-}22H]^{-7}$ at 2396.12 m/z). Peaks marked with "*" are the same product except missing the donor dye. Additional peaks in (c) and (d) are chemical background noise.

### S2.3. FRET efficiency calculation and spectral normalization

We estimate FRET efficiencies using the fluorescence ratiometric method, which is accurate when the excitation wavelength does not directly excite the acceptor dye and when all paired strands with a donor dye also have an acceptor dye.[4,5] These conditions are met by our selecting Alexa 488-Alexa 647 as the D-A dyes; using an excitation wavelength of 450 nm, which does not directly excite the acceptor; and using purified samples in which all donor labeled D-A pairs also contain an acceptor dye. This method finds the FRET efficiency, E, using the ratio of emission intensities from the donor and acceptor dye channels for each sample. Mathematically this is represented as:

$$E = \frac{1}{(\frac{\phi_A I_D}{\phi_D I_A}+1)} \qquad \text{(S2)}$$

Here, $E$ is the FRET efficiency, $\phi_A$ and $\phi_D$ are the acceptor and donor dye quantum yields, respectively, and $I_A$ and $I_D$ are the acceptor and donor dye integrated emission intensities, respectively. We use the standard quantum yield values of 0.92 for the donor dye, Alexa 488, and 0.33 for the acceptor dye, Alexa 647. For the cytosine case (Fig. 2a, main text), we find FRET efficiencies of 0.64 $\pm$ 0.01 for **D3′•A3′** and 0.07 $\pm$ 0.01 for **D5′•A3′**, where the error bars reflect uncertainties in the baseline for the emission intensity. For the guanine case (Fig. 2b, main text), we find FRET efficiencies of 0.64 $\pm$ 0.01 for **D3′•A3′** and 0.12 $\pm$ 0.01 for **D5′•A3′**. In both cases the higher FRET efficiency for **D3′•A3′** than for **D5′•A3′** indicates parallel pairing by $Ag^+$, as discussed in the main text.

In Figure 2 of the main text the **D3′•A3′** (solid blue) and **D5′•A3′** (orange) fluorescence spectra are normalized so that the peak donor emission is (1-$E$), corresponding to a peak value of 1 for emission from the bare donor strand (dotted blue curve; $E$ = 0 due to the absence of an acceptor dye).

For the **D3′•A3′** duplexes, which hold the D and A dyes on the same end of the duplex, the FRET efficiencies and the relation $E = 1/(1+(R/R_0)^6$ give a separation of 4.7 nm between the dyes, consistent with the duplex width plus the additional dye separation that arises from the thymine extensions and alkyl dye linkers. For the **D5′•A3′** duplexes, which hold the dyes on opposite duplex ends, the FRET efficiencies give separations of 7.2 nm for the guanine case and 8.0 nm for the cytosine case. A quantitative calculation of the lengths of the $Ag^+$-paired segments between the dyes is precluded by the flexibility of the T extensions and alkyl linkers to the dye labels. If we make the simplifying assumption that the duplex width is 2.2 nm, as for B-DNA, and that the T extensions plus alkyl linkers extend from the duplex at roughly 45º angles, the estimated length of the linear $[dG_{15}]_2$•$Ag_{15}$ duplex is roughly 5 nm. The corresponding average Ag—Ag distance of roughly 0.3 nm is in reasonable agreement with the modeling (main text and Section S4). For the cytosine case, the same approach gives a somewhat smaller average Ag—Ag distance; however the dye separation may be reduced if the cytosine duplex is less straight than the guanine duplex (Fig. 3D,E in main text).

## Section S3. IM-MS Ion moblity mass spectrometry

### S3.1. Experimental methods and data processing

The drift tube was operated in Helium at a pressure of 3.89 ± 0.01 Torr. Step-field experiments (five drift tube voltages for each samples) were performed to determine the CCS.

The arrival time distributions (ATDs) for each charge state of the complexes were fitted with 1 or 2 gaussian peaks using OriginPro 2016, to determine the arrival time $t_A$ of the center of the peak.

The arrival time $t_A$ is related to $\Delta V$ (voltage difference between the entrance and the exit of the drift tube region) by:

$$t_A = \frac{L^2}{K_0}\frac{T_0 p}{p_0 T} \cdot \left(\frac{1}{\Delta V}\right) + t_0 \qquad \text{(S3)}$$

$t_0$ is the time spent outside the drift tube region and before detection. A graph of $t_A$ vs. $1/\Delta V$ provides $K_0$ from the slope and $t_0$ as the intercept. The drift tube length is $L$ = 78.1 cm, the temperature is measured accurately by a thermocouple (here, $T$ = 297 ± 1 K), and the pressure is measured by a capacitance gauge (p = 3.89 ± 0.01 Torr). The CCS is then determined using Equation S4:

$$CCS = \frac{3ze}{16N_0} \cdot \sqrt{\frac{2\pi}{\mu k_B T}} \cdot \frac{1}{K_0} \qquad \text{(S4)}$$

The reconstruction of the experimental CCS distributions from the arrival time distributions and creation of the violin plots are performed as described elsewhere.[6] The x-axis of the graph is the number of bases and the y-axis is the mirrored CCS distribution. The violin plot shows simultaneously the difference in CCS values as a function of the charge state and the width of the CCS values distribution. For a given stoichiometry, the intensity ratios of the different charge states are taken into account with the size of the colored violin plots.

**S3.2. Supplementary results**

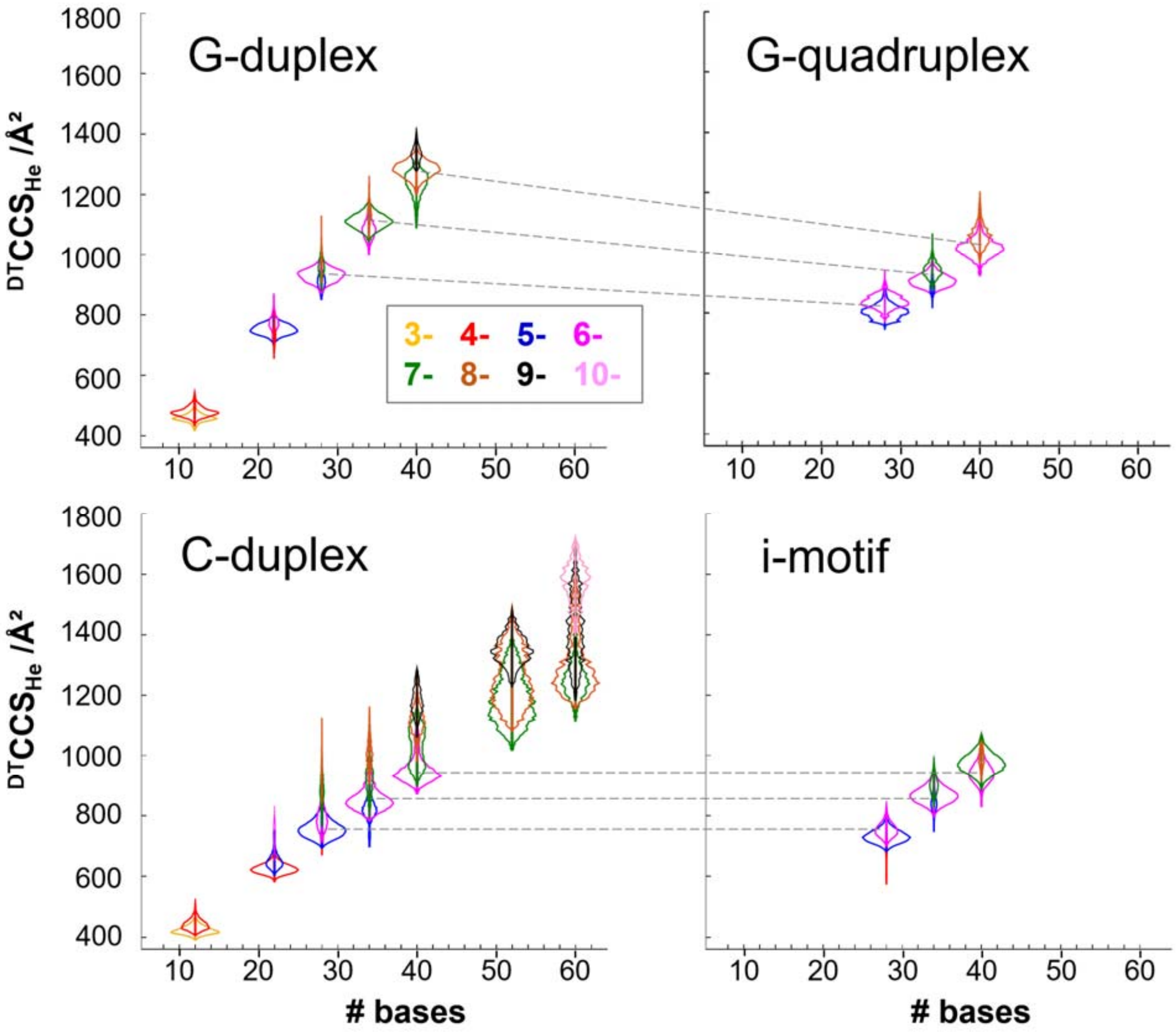

**Figure S12:** Comparison of the collision cross section distributions of G-duplexes $[dG_n]_2{\bullet}Ag_n$ and bimolecular G-quadruplexes $[dTG_nT]_2{\bullet}(NH_4^+)_i$, and of C-duplexes $[dC_n]_2{\bullet}Ag_n$ and i-motifs $[dC_n]_2$. The dashed lines indicate architectures containing the same number of bases. G-quadruplexes are more compact than G-duplexes containing the same number of bases. This is expected, given that the G-quadruplexes are tetra-helical, and that the G-duplexes are proposed to be double-helical. Thus, the number of G-quartets in the quadruplexes is lower than the number of G–$Ag^I$–G base pairs, so the overall topology is more globular. In the case of C-duplexes and i-motifs, similar CCS values are found. This is also expected because both contain base pairs (C–$Ag^I$–C for the C-duplex, and intercalated C–$H^+$–C for the i-motif) and thus the aspect ratios may be similar.

**Section S4. Molecular Modelling and Theoretical calculations**

**S4.1. Stucture generation and methods**

The structure of parallel right-handed poly(A) RNA (PDB: 4JRD)[7] was used as a starting material. It has an adequate spacing for two purines and the backbone was modified from RNA to DNA. The G–$Ag^{I}$–G base pair geometry was taken from the PBD structure 5XJZ,[8] and then fit in the parallel backbone.

Geometry optimizations and the molecular dynamics were performed using Gaussian 16 rev. A.03 software.[9] The $[dG_6]_2$•$Ag_6$ duplex and $[dC_6]_2$•$Ag_6$ duplex were optimized using DFT with M06-2X functional[10] including the dispersion correction GD3.[11] The 4-31G* basis set was used for the atoms C, H, O, N, P and the LanL2DZ basis set together with the associated effective core potential was used for the Ag atoms. Different basis sets (with the exception of the Ag atoms) (3-21G, 4-31G* and 6-31G*) and functionals (M06-2X, B3LYP) were tested and gives similar structures of the complexes (Figure S13).

Born-Oppenheimer molecular dynamics was performed (using gradient only) with a time step of 0.2 fs at the DFT level (M06-2X, 4-31G* or 3-21G*, GD3). 240 fs trajectory was produced for the $[dG_6]_2$•$Ag_6$ duplex and 500 fs trajectory for the $[dC_6]_2$•$Ag_6$ duplex. Figure S16 shows the adjacent Ag-Ag distances extracted every 2 fs for the $[dG_6]_2$•$Ag_6$ and $[dC_6]_2$•$Ag_6$ duplexes.

**S4.2. Calculations of the theoretical collisional cross section $^{TM}CCS_{He}$**

The $^{TM}CCS_{He}$ were calculated with the mobcal code[12,13] using the trajectory method (TM). The silver atoms were replaced by nitrogen (Silver is not parametrized in the TM model). The $^{TM}CCS_{He}$ were calculated on structures of the complexes extracted every 2 fs from the BOMD trajectories. Histogram plots of the CCSs were generated using a number of bins of 40.

### S4.3. Supplementary figures

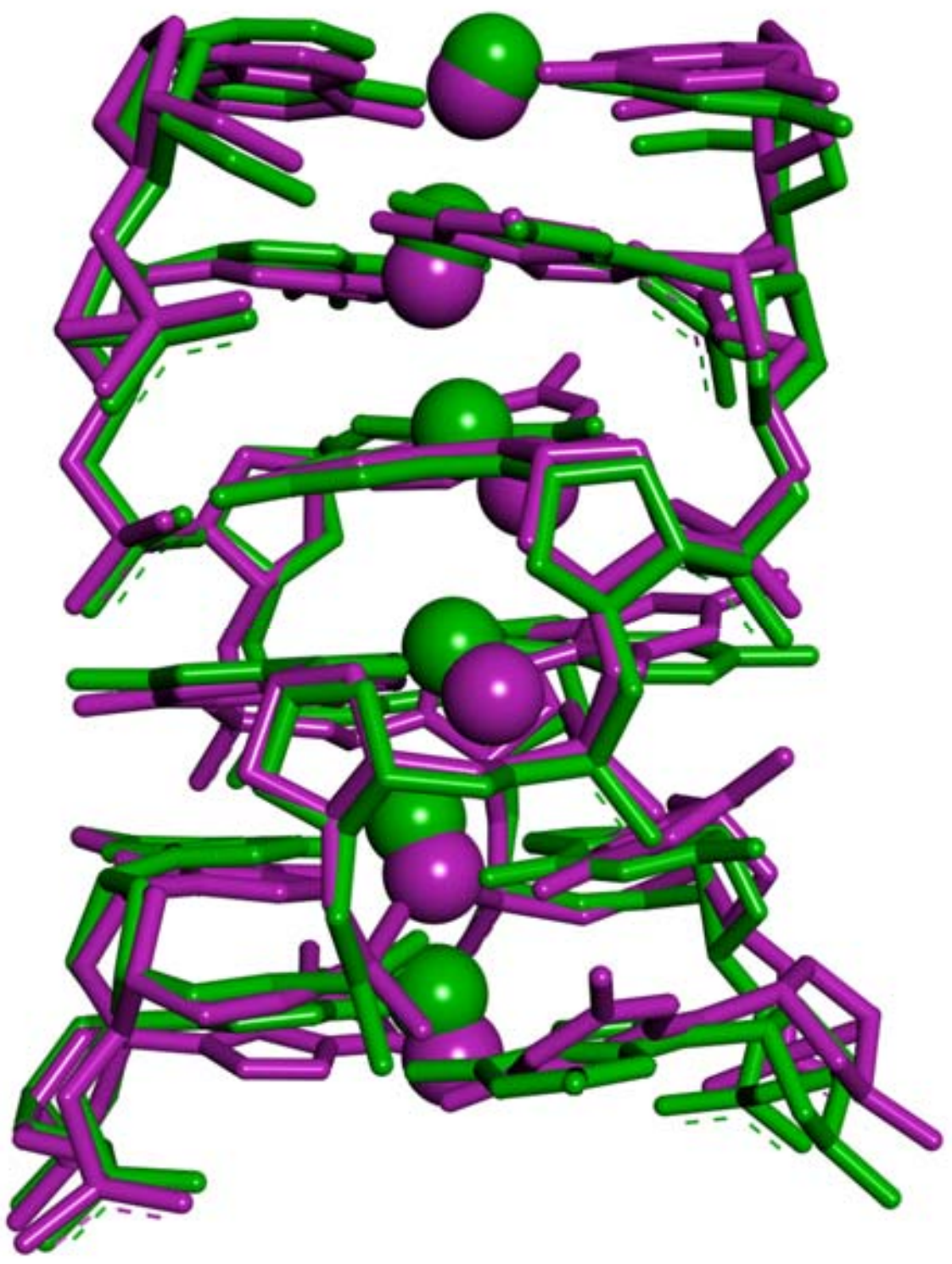

**Figure S13:** Comparison of the DFT optimized (4-31G* + GD3) structures of the $[(dG_6)_2{\bullet}Ag_6]^{4-}$ using B3LYP (violet) and M06-2X (green). Overall shape is the same ($RMSD_{all\ atoms}$ = 0.61 Å). The calculated CCS values are comparable.

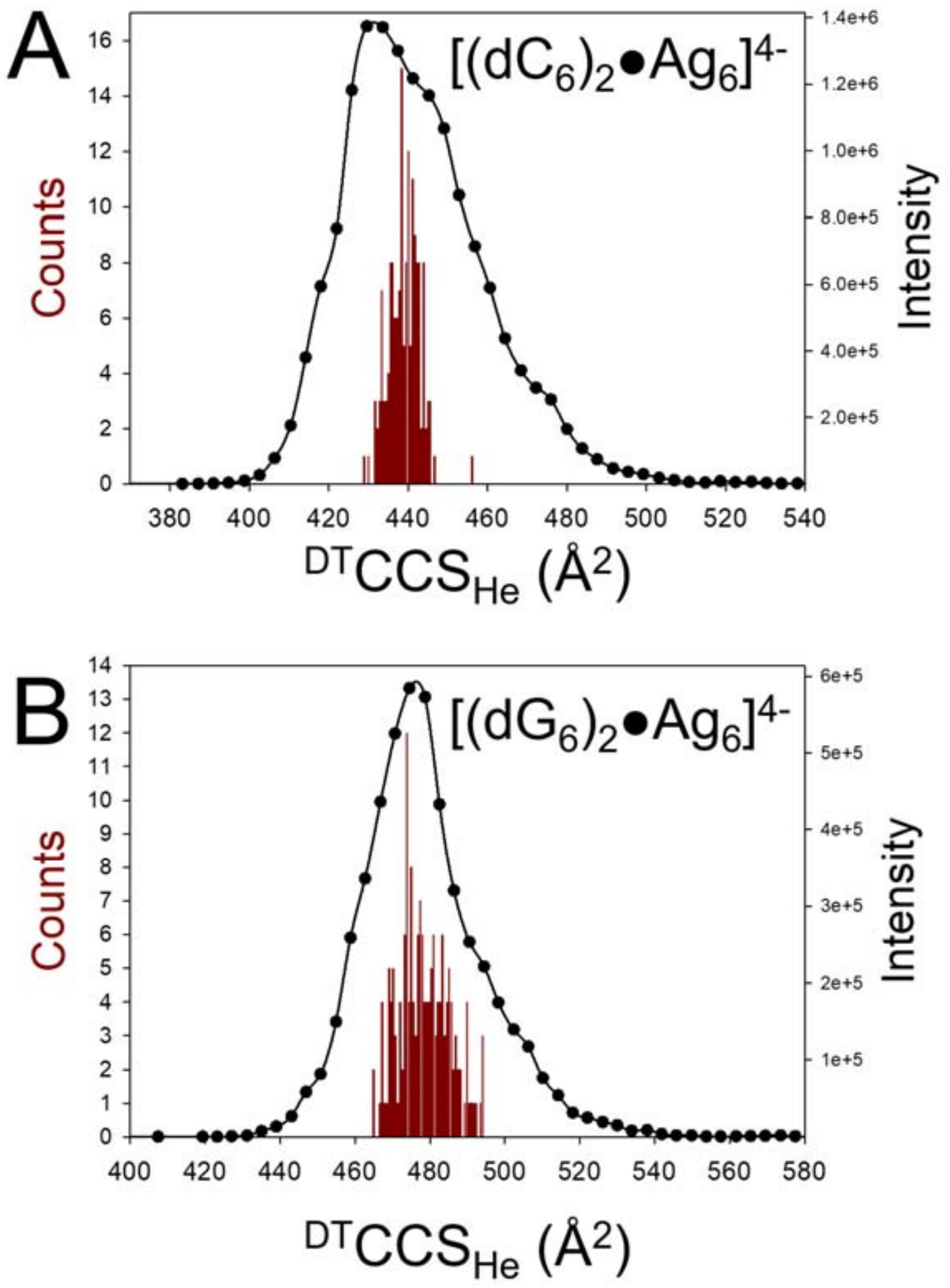

**Figure S14:** Comparison between the experimental (black dots) and theoretical (red histograms) collision cross sections for $[(dC_6)_2\bullet Ag_6]^{4-}$ (A) and $[(dG_6)_2\bullet Ag_6]^{4-}$ (B). The histograms are constructed with 40 bins.

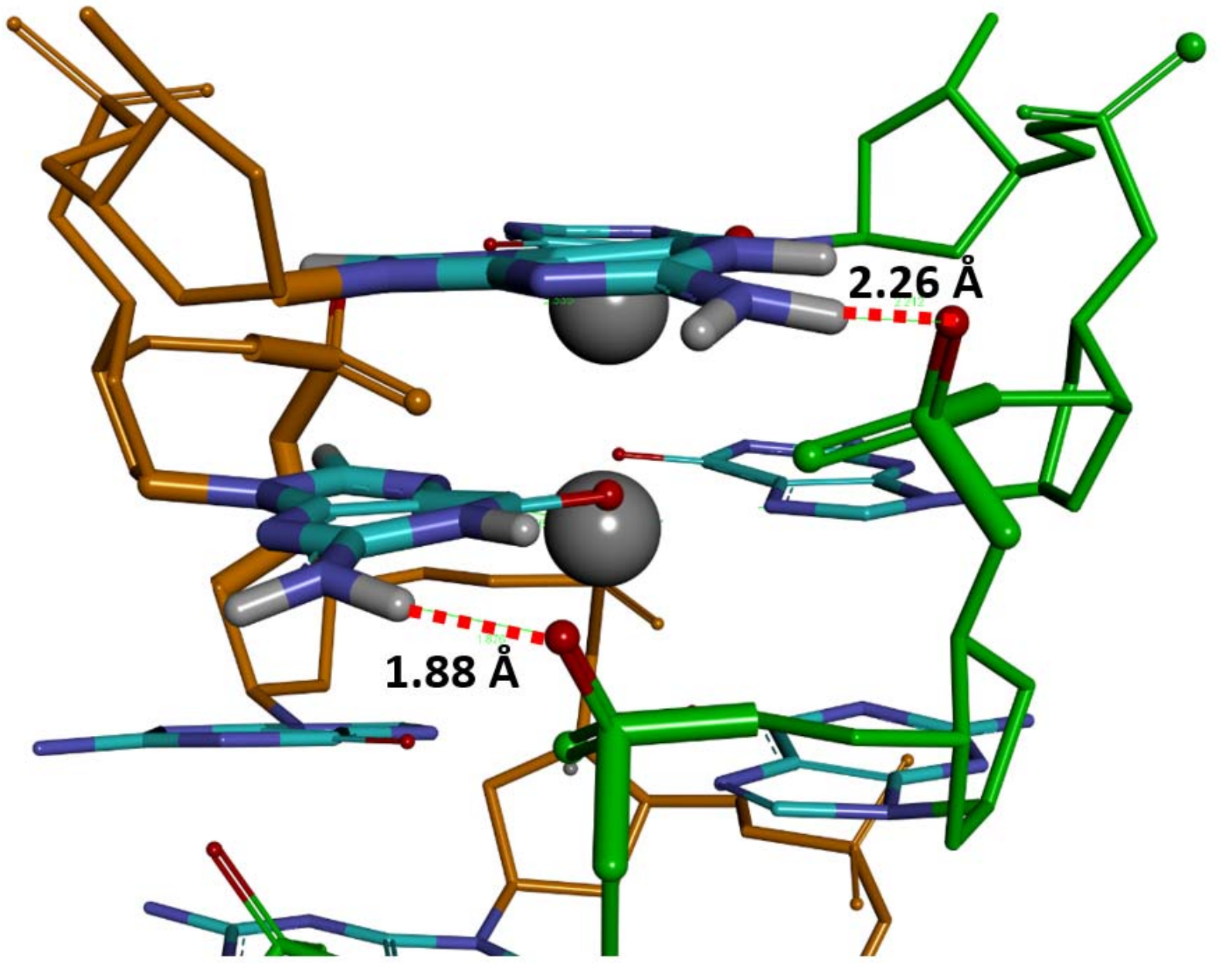

**Figure S15:** Hydrogen bonding between the guanine NH2 and the phosphate oxygens in the M06-2X-optimized structure of $[(dG_6)_2{\bullet}Ag_6]^{4-}$
.

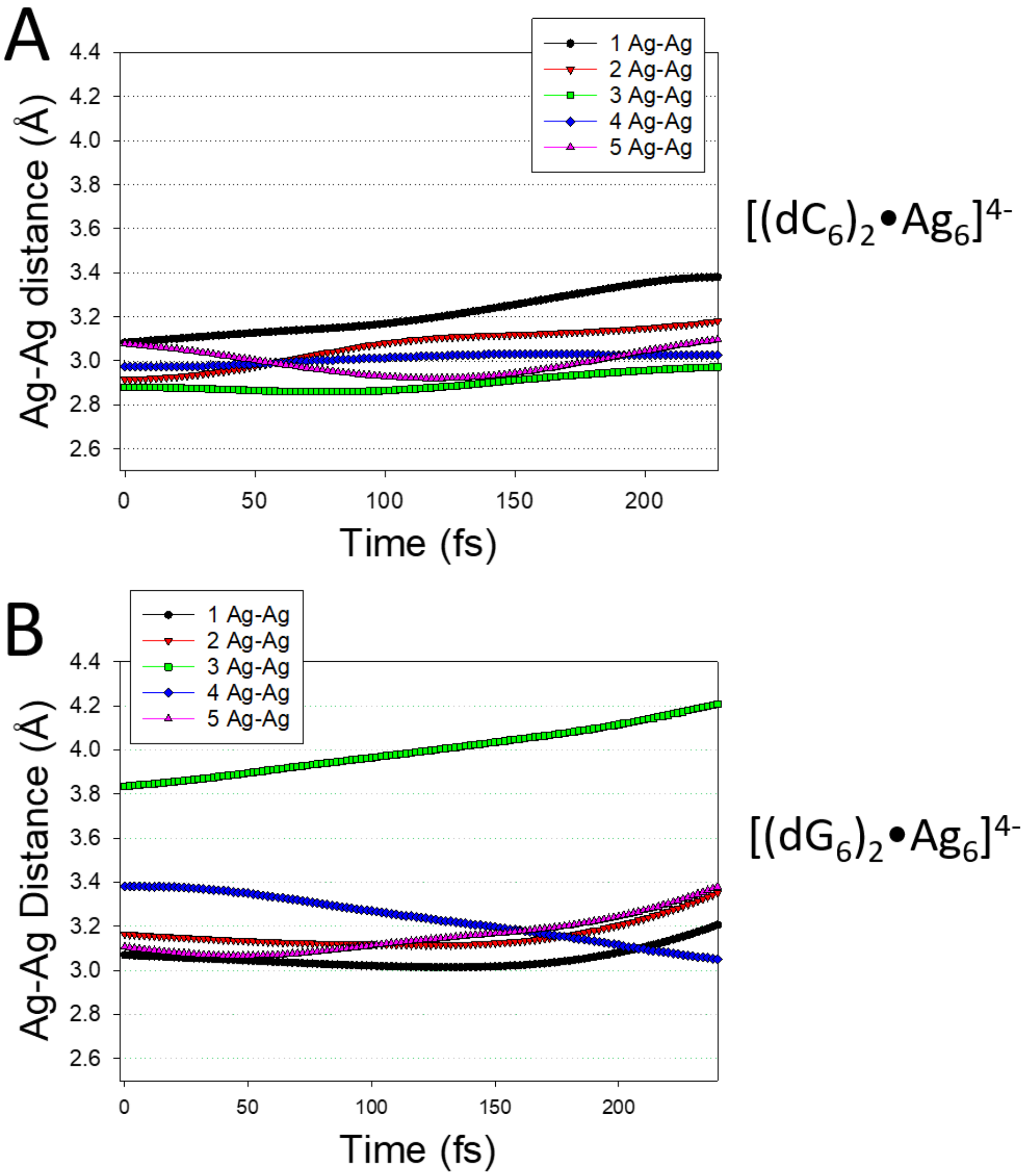

**Figure S16:** Distance between neighboring silver atoms monitored during the ab initio MD. "1 Ag-Ag" means the distance between the lowest and next lowest silver cation, starting from the 3´end of the duplex. (A) $[(dC_6)_2{\bullet}Ag_6]^{4-}$ over 500 fs. All Ag—Ag distances comply with the definition of argentophilic bonds. (B) $[(dG_6)_2{\bullet}Ag_6]^{4-}$ over 240 fs. The silver atoms are distributed in two groups of three, as evidenced by the larger separation ("3 Ag-Ag") between the 3rd and 4th silver cations in the duplex.

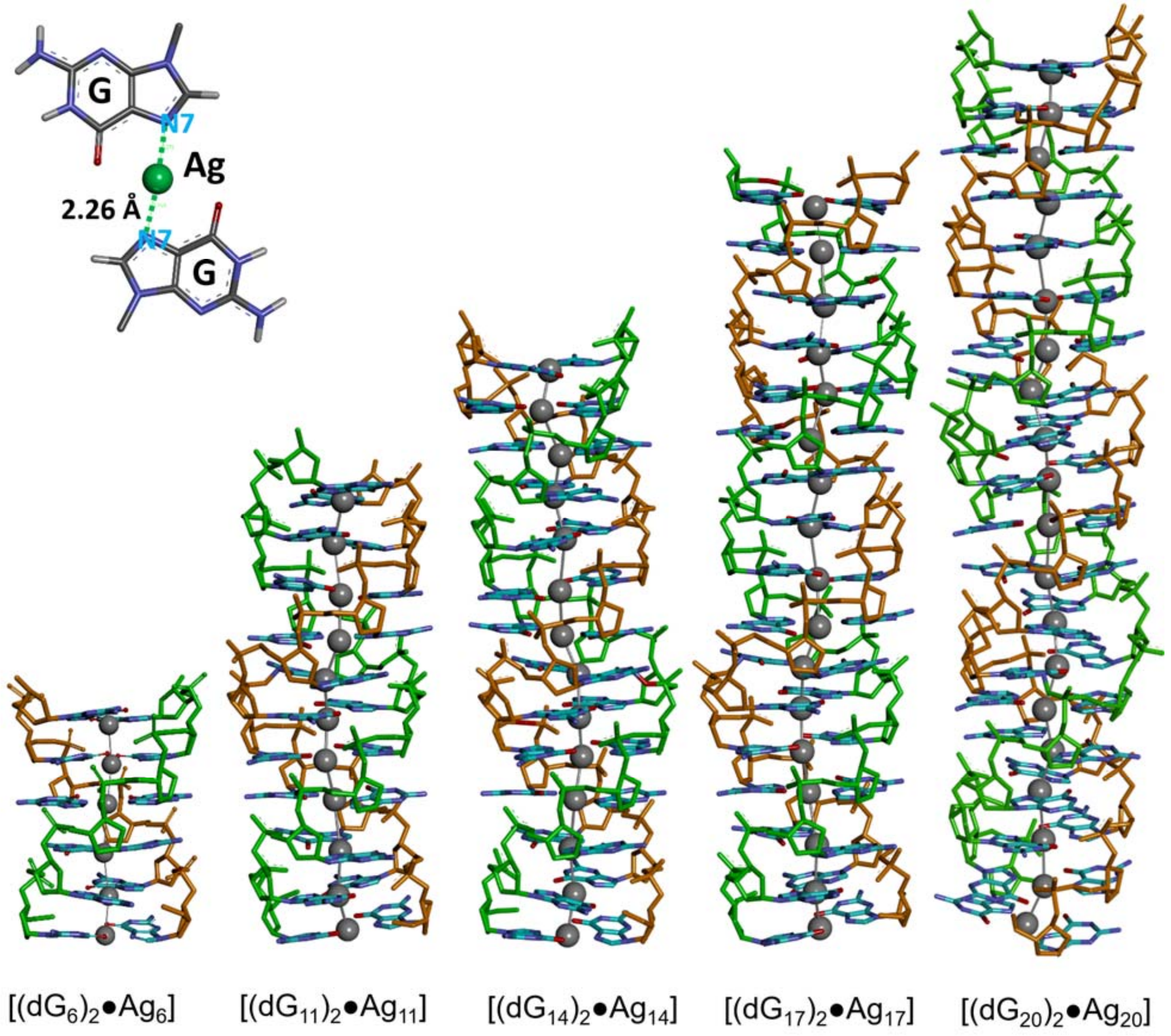

**Figure S17:** Structures of the $[(dG_n)_2{\bullet}Ag_n]$complexes generated by concatenating the DFT optimized structure of $[(dG_6)_2{\bullet}Ag_6]^{4-}$. These structures were used to calculate theoretical collision cross sections ($^{TM}CCS_{He}$) for the trend line of Figure 3A. The hydrogen atoms are hidden for clarity. The two backbone strands are colored in green and brown, guanines are in blue, and silver atoms are in grey. Inset: postulated configuration of the G-$Ag^I$-G base pairs.

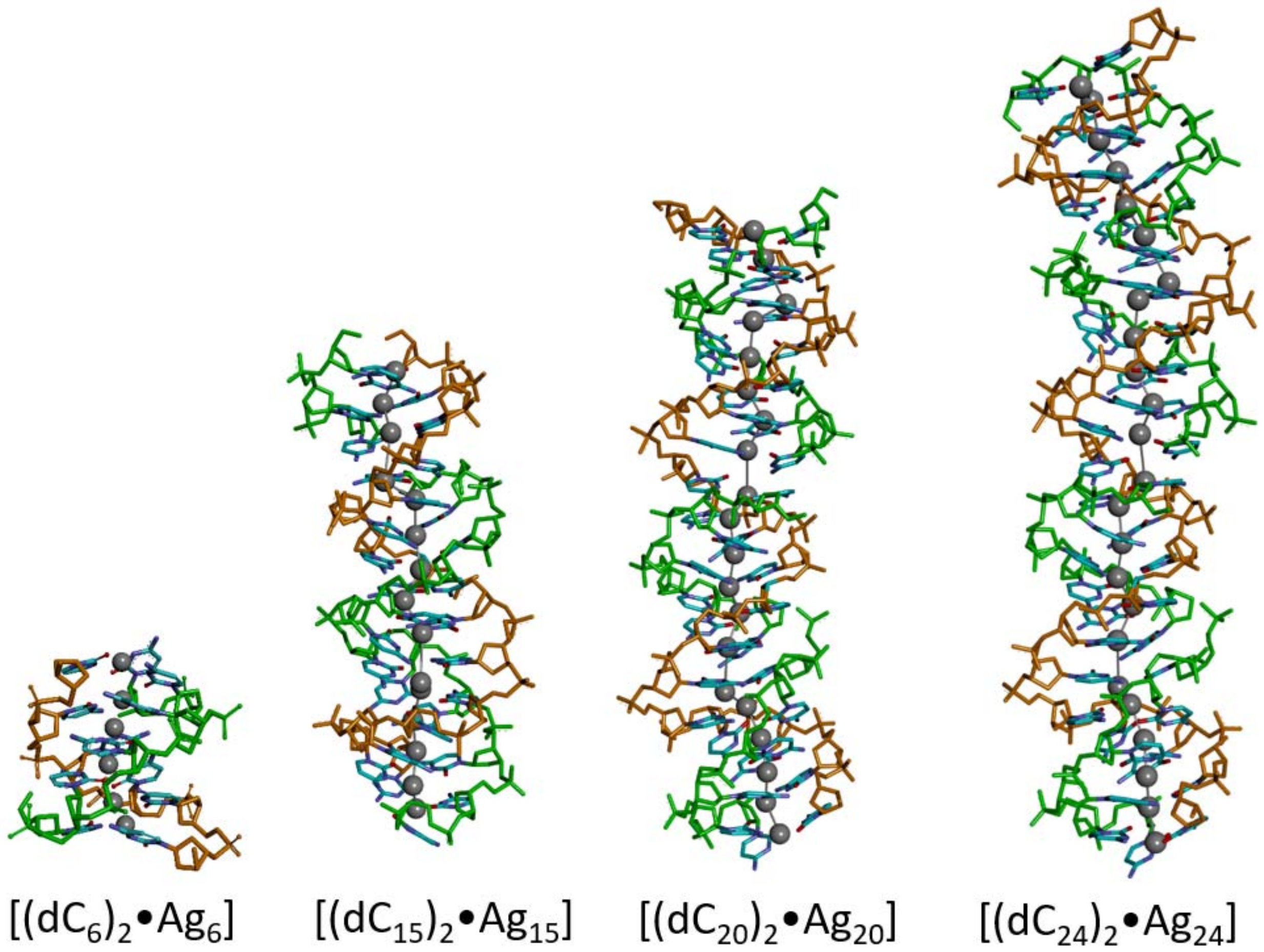

**Figure S18:** Structures of the $[(dC_n)_2\bullet Ag_n]$ complexes generated by concatenating the DFT optimized structure of $[(dC_6)_2\bullet Ag_6]^{4-}$. These structures were used to calculate theoretical collision cross sections ($^{TM}CCS_{He}$) for the trend line of Figure 3A. The hydrogen atoms are hidden for clarity. The two backbone strands are colored in green and brown, guanines are in blue, and silver atoms are in grey.

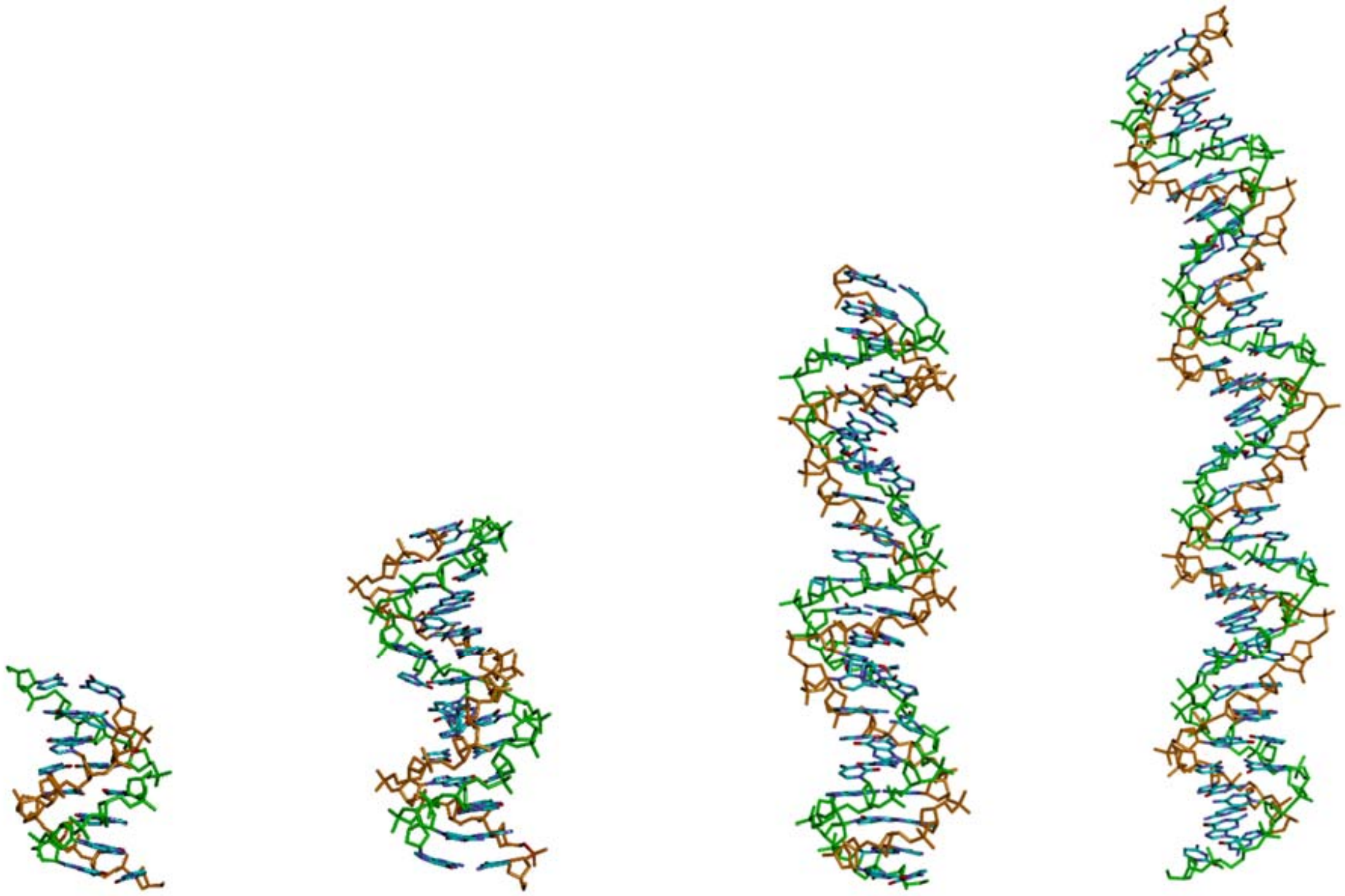

**Figure S19:** Structures of the gas-phase duplexes [d(CG)$_n$]$_2$, generated by concatenating the DFT optimized structure of [d(CG)$_4$]$_2^{4-}$. These structures were used to calculate theoretical collision cross sections ($^{TM}CCS_{He}$) for the trend line of Figure 3C.

**Section S5. Supplementary references**